\documentclass[a4paper,11pt]{article}
\pdfoutput=1
\usepackage{jheppub}

\usepackage{textcase}
\usepackage{graphicx,epsfig,psfrag,amssymb,hyperref}
\usepackage{multirow}
\usepackage{color,graphicx,epsfig,psfrag,amsmath,empheq}
\usepackage{bm}
\usepackage{mathrsfs,amsfonts,soul,color}
\usepackage[caption=false]{subfig}
\usepackage{slashed}

\usepackage{placeins}

\usepackage[utf8]{inputenc}

\newcommand{\abs}[1]{\left| #1\right|}
\newcommand{\avg}[1]{\left\langle #1\right\rangle}

\newcommand{\cL}{\mathcal{L}}

\newcommand{\cO}{\mathcal{O}}

\newcommand{\Ep}{E^{\prime}}

\newcommand{\keV}{{\rm keV}}
\newcommand{\MeV}{{\rm MeV}}
\newcommand{\TeV}{{\rm TeV}}
\newcommand{\GeV}{{\rm GeV}}
\newcommand{\eV}{{\rm eV}}

\newcommand{\kinmix}{\epsilon}
\newcommand{\mchik}{m_{\chi{\keV}}}
\newcommand{\mET}{m_{E{\TeV}}}
\newcommand{\lamFS}{\lambda_{\rm FS}}

\newcommand{\be}{\begin{equation}}
\newcommand{\ee}{\end{equation}}
\newcommand{\bea}{\begin{eqnarray}}
\newcommand{\eea}{\end{eqnarray}}

\title{Stealth decaying spin-1 dark matter}
\preprint{LAPTH-041/20}

\author[a]{C\'edric Delaunay,}
\author[b]{Teng Ma,}
\author[b]{and Yotam Soreq}

\affiliation[a]{LAPTh, CNRS -- USMB, BP 110 Annecy-le-Vieux, F-74941 Annecy, France}
\affiliation[b]{Physics Department, Technion -- Israel Institute of Technology, Haifa 3200003, Israel}

\emailAdd{cedric.delaunay@lapth.cnrs.fr}
\emailAdd{t.ma@campus.technion.ac.il}
\emailAdd{soreqy@physics.technion.ac.il}

\abstract{
We consider models of decaying spin-1 dark matter whose dominant coupling to the standard model sector is through a dark-Higgs Yukawa portal connecting a TeV-scale vector-like lepton to the standard model (right-handed) electron. Below the electron-positron threshold, dark matter has very slow, loop-suppressed decays to photons and (electron) neutrinos, and is stable on cosmological time-scale for sufficiently small gauge coupling values. Its relic abundance is set by in-equilibrium dark lepton decays, through the freeze-in mechanism. 
We show that this model accommodates the observed dark matter abundance for natural values of its parameters and a dark matter mass in the $\sim 5\,\keV$ to $1\,\MeV$ range, while evading constraints from direct detection, indirect detection, stellar cooling and cosmology. We also consider the possibility of a nonzero gauge kinetic mixing with the standard model hypercharge field, which is found to yield a mild impact on the model's phenomenology.
}
\begin{document}
\maketitle

%%%%%%%%%%%%%%%%%%%%%%%%%%%%%%%%%%%%%%%%%%%%%%%%%%%%%%%%%%%%%%%%%%%%%%
\section{Introduction}
\label{sec:Intro}
%%%%%%%%%%%%%%%%%%%%%%%%%%%%%%%%%%%%%%%%%%%%%%%%%%%%%%%%%%%%%%%%%%%%%%

Dark matter~(DM) is perhaps the most fascinating puzzle in high energy physics today.
The existence of DM is supported by different observations from galactic to cosmic scales. However very little is known about its nature. For instance, it is not even clear that DM is made of fundamental particles. If so, most properties of particle DM, like its mass, spin and relic abundance generation mechanism, are unknown.
Moreover, the observational evidence for DM is all based on gravity, and the existence of other types of interaction with the Standard Model~(SM) fields remain to be discovered.\\

Non-gravitational interactions of the DM are often advocated to explain the production of its relic abundance in the early universe. 
This assumption triggered hope to directly detect DM in terrestrial experiments~\cite{PhysRevD.31.3059} or to produce it at colliders. 
Despite a sustained effort over several decades, experiment has failed to discover theoretically well-motivated DM candidates, such as weakly interacting massive particles~(WIMPs)~\cite{Akerib:2016vxi, Cui:2017nnn,Aprile:2018dbl} and axion DM~\cite{Du:2018uak,Ouellet:2018beu}. 
This null result calls for either alternative candidates~\cite{Battaglieri:2017aum}  or alternative relic production mechanisms, and perhaps even questions the very existence of DM interactions beyond gravity~\cite{Garny:2015sjg,Garny:2017kha,Ema:2018ucl,Garny:2018grs}. 

One possible alternative scenario consists in realizing the DM as feebly interacting massive particles~(FIMPs) that are produced through the freeze-in mechanism~\cite{McDonald:2001vt,Hall:2009bx}. 
See Ref.~\cite{Bernal:2017kxu} for a recent review.   
Unlike the standard thermal scenario where the DM is in equilibrium with the SM until its annihilation rate is beaten by Hubble expansion leading to DM freeze-out, 
within the freeze-in scenario the DM is never in thermal equilibrium with the SM and is gradually produced from scattering or decay of SM particles. 
Despite its non thermal nature the freeze-in mechanism, similarly to the freeze-out one, is typically not sensitive to initial conditions in the very early Universe, with the known exception of UV-dominated freeze-in~\cite{Elahi:2014fsa}.  
Consequently, the relic abundance only depends on DM properties that can be probed in experiments. Also, FIMP may be related to leptogenesis~\cite{Falkowski:2017uya}.

Massive spin-1 particles are interesting DM candidates. 
In particular, for masses less than that of the electron-positron pair, new vectors can only decay to $3\gamma$ or $\nu\bar\nu$ with  typically highly suppressed decay widths, such that vector DM~(VDM) particles are easily stable on cosmological time scales without resorting to an ad-hoc stabilizing symmetry.  A similar feature is observed in models of scalar DM decaying to SM fields through the Higgs portal~\cite{Heeba:2018wtf,Fradette:2018hhl}.  A well-know VDM candidate enjoying this property is dark photon DM~\cite{Pospelov:2008jk,Redondo:2008ec,Bjorken:2009mm}, which only couples to the SM through kinetic mixing with the hypercharge field~\cite{Okun:1982xi,Galison:1983pa,Holdom:1985ag}. 
However, it is rather challenging to construct a viable model of dark photon DM below the MeV scale. 
For instance, a thermally produced relic abundance is excluded by direct detection searches~\cite{Redondo:2008ec,An:2014twa,Hochberg:2016sqx} and in strong tension with stellar cooling~\cite{Redondo:2008aa, An:2013yfc, Redondo:2013lna} and cosmological bounds~\cite{McDermott:2019lch}. 
Sufficient dark photon production is still achievable nonthermally. 
For instance, dark photons can be produced during inflation by a misalignment mechanism~\cite{Nelson:2011sf} or quantum fluctuations~\cite{Nelson:2011sf,Graham:2015rva}. 
Another possibility is dark photon production from an oscillating axion field~\cite{Agrawal:2018vin,Co:2018lka,Dror:2018pdh,Bastero-Gil:2018uel}. Finally, nonthermally produced massive spin-1 DM may also be related to neutrino masses and leptogenesis in the context of $B-L$ models~\cite{Choi:2020dec}.\\

In this work we explore a model of sub-MeV VDM resulting from the spontaneous breakdown of a very weakly gauged dark U(1)$_X$ group. 
Its relic abundance is created from the decay of a heavy dark fermion in thermal equilibrium with the SM. 
The dark fermion is a massive isospin singlet vector-like lepton at the TeV scale or higher, unit dark charge and hypercharge $-1$, which decays to VDM thanks to mass mixing with the SM leptons. 
There are two possible VDM production regimes, depending on the strength of the dark sector interactions with the SM. For feeble coupling values the VDM relic density is set by the freeze-in mechanism, whereas in the limit of couplings large enough to bring VDM in equilibrium with the thermal bath, VDM freezes out while being still relativistic once the dark fermion decouples. The latter case is reminiscent of the forbidden DM scenario~\cite{DAgnolo:2015ujb}, although in a limit where DM is much lighter than the mediator.   

This minimal setup has several interesting features. 
First of all, there is a parametric suppression of the VDM coupling to SM lepton pairs relative to its coupling to one SM and one dark lepton. 
This allows to saturate the observed DM relic abundance while evading all existing constraints. 
Then, the VDM-to-SM couplings needed for DM production are all technically natural, and can thus assume arbitralily small values without introducing fine-tuning. As a result, there is a large portion of parameter space where DM can be naturally produced to the observed level and yet remain invisible to all DM detection experiments.  Finally, in the relativistic freeze-out limit, the comoving VDM number density at freeze-out is independent of its mass and couplings, up to possible change of the number of entropy degrees of freedom during freeze-out. Hence, in this regime, the observed DM abundance requires  a hot DM with a $\sim 60\,$eV mass, which is in tension with large-scale structure~(LSS) formation~\cite{Colombi:1995ze,Bode:2000gq,Viel:2013fqw}.

The rest of the paper is organized as follows. 
We define our VDM model in Section~\ref{sec:Model}, calculate its relic abundance in Section~\ref{sec:FreezeIn} and present its DM-related phenomenology and collider signatures in Section~\ref{sec:signatures} . 
We summarize our conclusions in Section~\ref{sec:summary}.

%%%%%%%%%%%%%%%%%%%%%%%%%%%%%%%%%%%%%%%%%%%%%%%%%%%%%%%%%%%%%%%%%%%%%%
\section{Light dark matter from a feebly gauged dark U(1)$_X$ group}
\label{sec:Model}
%%%%%%%%%%%%%%%%%%%%%%%%%%%%%%%%%%%%%%%%%%%%%%%%%%%%%%%%%%%%%%%%%%%%%%

In this section, we present a minimal model of VDM with a sub-MeV mass.
Consider, in addition to the SM symmetries and fields, a U(1)$_X$ gauge group with a tiny gauge coupling $g_X\ll 1$ which is spontaneously broken by a dark Higgs $\phi$ of U(1)$_X$ charge unity. 
Consider also a dark vector-like fermion $E$ which has the same charge as $\phi$ under U(1)$_X$ and carries hypercharge $-1$. 
We assume all SM fields to be U(1)$_X$ neutral.

The most general renormalizable Lagrangian is then,
\begin{align}
	\label{eq:Lall}
	\cL
= 	\cL_{\rm SM} + \cL_{\rm dark} + \cL_{\rm portal} \, ,
\end{align}
where $\cL_{\rm SM}$ is the SM Lagrangian and
\begin{align}
	\cL_{\rm dark}
=&	-\frac{1}{4} \chi_{\mu\nu} \chi^{\mu\nu}
	+\overline{E}\left( i\slashed{D} -M_E\right) E
	+(D_\mu \phi)^\dagger D^\mu \phi - V(\phi) \, ,  \\
	\cL_{\rm portal}
=&	-\lambda_{\phi H} \abs{\phi}^2 \abs{H}^2
	- \frac{\kinmix}{2} \chi_{\mu\nu} B^{\mu\nu}
	- \left( x_E \overline{E}_L e_R \,\phi+ {\rm h.c.} \right) \, ,
\end{align}
where $H$ is the SM-Higgs field, $B_{\mu}$ is the hypercharge gauge field, with gauge coupling $g_Y$, and $e_R$ denotes the right-handed~(RH) SM leptons.

We assume that $V(\phi)=\lambda_\phi(|\phi|^2-w^2/2)^2$ is unstable at the origin and has its global minimum at $w > 0$. 
In unitary gauge $\phi = (\varphi + w)/\sqrt{2}$ where $\varphi$ denotes a real scalar fluctuation around the non zero vacuum expectation value~(VEV) $w$. 
This spontaneous breakdown induces a mass for the U(1)$_X$ gauge field of  $m_\chi = g_X w\,$. 
Note that a natural value of the dark Higgs VEV is (assuming the quartic coupling dominates the quantum correction to the $\phi$ mass squared) $w\sim \lambda_\phi^{1/2} \Lambda/4\pi$ where $\Lambda$ is the cutoff scale of the model. 
Hence, a moderately large cutoff of, for instance, $10^5\,\TeV$ together with a vector mass $m_\chi<2m_e$ implies a tiny value of the U(1)$_X$ gauge coupling of
\begin{align}
	\label{gXbound}
	g_X 
	\sim 
	10^{-13}\, \frac{\mchik}{\lambda_\phi^{1/2}} \left(\frac{10^5\,{\TeV}}{\Lambda}\right)\,,
\end{align}  
where $\mchik$ is the VDM mass in units of $\keV$.
For fixed $m_\chi$ and $\lambda_\phi$, larger values of  $g_X$ are possible at the expense of fine-tuning the VEV $w$ below its natural value (or lowering the cutoff scale), while smaller values can be obtained by raising the scale $\Lambda$. Another implication of naturalness is that the scalar fluctuation $\varphi$ is very heavy $m_\varphi\sim\lambda_\phi^{1/2}\Lambda/4\pi \sim 8 \times 10^4 \,\lambda_\phi^{1/2} (\Lambda/10^5\,\TeV)\,\TeV$, unless the quartic coupling is very small. Henceforth, we assume $\lambda_\phi\sim\cO(1)$. The vector-like fermion mass $M_E$ can be anywhere between a few hundred GeV and $\Lambda$, lighter masses being in tension with collider constraints. In the subsequent sections, we will focus on $M_E$ in the TeV range for definiteness.\\

The operators in $\cL_{\rm portal}$ are portal interactions linking the dark sector states to the SM. 
The first two terms are the Higgs and kinetic mixing portals, respectively, while the last one is a leptonic Yukawa portal. 
While the Higgs portal yields a rich phenomenology when the scalar $\varphi$ is light, see {\it e.g.} Ref.~\cite{Arcadi:2019lka} for a review, in our case the scalar fluctuation is close to the cutoff scale and totally decoupled. 
Therefore, for simplicity, we set $\lambda_{\phi H}=0$. 

The kinetic mixing however is not always negligible. Even if set to zero at the classical level, it is generated at the one-loop level from states carrying both hypercharge and U(1)$_X$ charge. 
There is only one such fermion in our model, giving~\cite{Holdom:1985ag}
\begin{align}
	\epsilon_{\rm loop} 
= 	\frac{g_Xg_Y}{6\pi^2}\log\left( \frac{M_E}{\Lambda} \right)\, ,
\end{align} 
which is logarithmically divergent and thus sensitive to an unknown UV contribution. The loop contribution to the kinetic mixing could be made calculable by adding extra fields.
For example, consider an additional heavy vector-like fermion $E^\prime$ of mass $M_\Ep$, hypercharge $-1$ and U(1)$_X$ charge opposite to that of $E$. 
Then, the total one-loop contribution is finite with $\kinmix_{\rm loop} = g_X g_Y / (6\pi^2)\log\left(M_E/M_{\Ep}\right)$. 
The kinetic mixing even vanishes in the limit of degenerate fermions ($M_E=M_{\Ep}$), which may result from an approximate $Z_2$ symmetry. 
In principle, this would require to include an additional portal operator like $- x_{\Ep} \overline{\Ep}_L e_R \,\phi^\dagger+ h.c.$ which would not change the phenomenology. 
Therefore, in this case one could set $x_{\Ep}=0$ for simplicity, so the only effect of $\Ep$ would be to regulate the kinetic mixing.\footnote{ 
This assumption explicitly breaks the $Z_2$ symmetry and radiatively lifts the mass degeneracy, thus, reintroducing a nonzero $\kinmix_{\rm loop}$. 
However, we will consider a regime where $x_E\ll 1$ so that this breaking is small and negligible in practice.} To summarize $\kinmix$ is essentially a free parameter of the model, whose precise value depends on UV physics. In order to get a feeling of this freedom on the model's phenomenology, we investigate two representative cases in the following, where $\kinmix=0$  and $\kinmix=g_Xg_Y/(6\pi^2)$.\\

After U(1)$_X$ breaking, $E$ mixes with the SM electron. 
The corresponding mass matrix is diagonalized by the mixings angles   
\begin{align}
	\tan\left(2\theta_R\right)
&=	\frac{2\sqrt{2}M_E \, x_E w }{ 2M^2_E - (y_e v)^2 - (x_E w)^2 }  \,,
\end{align}
between $e_R$ and $E_R$, and
\begin{align}
	\tan\left(2\theta_L\right)
&= 	\frac{2y_e v  \,  x_E  w}{ 2M^2_E - (y_e v)^2 +(x_E w)^2 }  \, ,
\end{align}
between $e_L$ and $E_L$,
where $v\approx246\,\GeV$ is the SM Higgs VEV and $y_e$ is the electron Yukawa.
Its eigenvalues, denoted $m_e$ and $m_E$ (with $m_e<m_E$), obey
\begin{align}
	m_e m_E = y_e v M_E/\sqrt{2} \,,\quad
	m_e^2 + m_E^2 = M_E^2 + \left[ (y_e v)^2 + (x_E w)^2\right] /2 \,,
\end{align}
where $m_e\approx 511\,$keV is identified with the physical electron mass.
In the limit of $M_E\gg y_e v$ and $x_E w$ that we envisage here, we have approximately
$\theta_R\approx	 x_E w /(\sqrt{2} M_E)$ and  $\theta_L \approx
	 y_e v  \theta_R/( \sqrt{2} M_E)\ll \theta_R$
and the physical masses are corrected at $\cO(\theta_R^2)$ relative to the unmixed case as
$m_e\approx y_e v/\sqrt{2}( 1- \theta_R^2/2 )$ and	$m_E \approx M_E( 1+ \theta_R^2/2)$.

The mass mixing allows the dark vector $\chi_\mu$ to interact with the electron through the Lagrangian ($e$ and $E$ now denoting mass eigenstates)
\begin{align}
	\cL_{\rm int}
=	j_X^\mu \chi_\mu\,,
\end{align}
\begin{align}
j_X^\mu= \bar e\gamma^{\mu} (g_{\chi ee}^V +g_{\chi ee}^A \gamma_5) e
	+\bar E\gamma^{\mu} (g_{\chi EE}^V+g_{\chi EE}^A \gamma_5)E
	+ [\bar E\gamma^{\mu}(g_{\chi Ee}^V+g_{\chi Ee}^A\gamma_5)e +{\rm h.c.}]\,,
\end{align}
where $g^{V,A}\equiv (g^R\pm g^L)/2$ and  (at linear order in $\kinmix$, neglecting the subleading mixing contribution with the $Z$~\cite{Cassel:2009pu,Cline:2014dwa})
\begin{align}
	\label{eq:gchiff}
	g_{\chi ee}^{L,R} 
	\approx 
	g_Xs_{L,R}^2+\kinmix ec_W\,,\quad 
	g_{\chi Ee}^{L,R}
	\approx 
	g_Xc_{L,R} s_{L,R}\,,\quad 
	g_{\chi EE}^{L,R}
	\approx 
	g_Xc_{L,R}^2+\kinmix e c_W\,,
\end{align}
%*
 where 
$c_W\,(s_W) \equiv \cos \theta_W\,(\sin \theta_W)$, $\theta_W$ is the weak mixing angle 
and $c_{L,R}\equiv\cos\theta_{L,R}$, $s_{L,R}\equiv \sin\theta_{L,R}$.
In the absence of kinetic mixing, 
the LH couplings are suppressed by a factor of $\cO(m_e/m_E)$ relative to the RH ones and $g^R_{\chi ee}/g^R_{\chi Ee}\approx\theta_R\ll1$. 
This implies in particular that VDM production from annihilations of electron pairs will be parametrically suppressed, and typically negligible, compared to that from in-equilibrium $E$ decays which is controlled by the $E\to \chi e$ partial width given by
\begin{align}
	\label{E2eXwidth}
	\Gamma_{E \to \chi e} 
	\simeq
	\frac{g_X^2\theta_R^2m_E^3}{32\pi m_\chi^2} \simeq \frac{x_E^2 m_E}{64\pi}\,,
\end{align}
to leading order in $\theta_R$. 
Note that the enhancement factor $(m_E/m_\chi)^2$ arising from the longitudinal polarization of $\chi_\mu$ cancels out with a similar factor from the mixing angle, signaling that $E$ decays actually to the Goldstone boson of the broken U(1)$_X$ symmetry through the Yukawa portal operator. Electroweak (EW) decays $E\to Ze$ and $E\to  W\nu$ are also possible, with the following widths (again, to leading order in $\theta_R$)
\begin{align}
	\label{E2EWwidth}
	\Gamma_{E\to Ze}
	\simeq \frac{g_Z^2\theta_R^2m_Em_e^2}{128\pi m_Z^2}\left(1+4\kinmix^2\frac{g_X^2m_E^2}{g_Z^2m_e^2}\right)\,,\quad 
	\Gamma_{E\to W\nu} 
	\simeq \frac{g^2\theta_R^2m_Em_e^2}{64\pi m_W^2}\,,
\end{align} 
where $g$ is the SU(2)$_L$ gauge coupling, $g_Z\equiv g/c_W$, and assuming the large $m_E$ limit. Note that, in the absence of kinetic mixing, the decay to $Ze$ is mediated only by the LH mixing angle $\theta_L \simeq m_e\theta_R/m_E$. \\

The dark vector, $\chi_\mu$, is not stabilized by any symmetry and decays. 
For $m_\chi < 2m_e$ the leading decay channels are in three-photons, the two-photon decay being forbidden by the Yang theorem~\cite{PhysRev.77.242}, and $\nu\bar\nu$ final states. In the Euler-Heisenberg effective Lagrangian limit~\cite{Heisenberg:1935qt} the  $3\gamma$ decay rate is~\cite{Pospelov:2008jk,Redondo:2008ec} 
\begin{align}
	\label{eq:Gamma3gamma}
	\Gamma_{\chi\to 3\gamma}
	\simeq
	\frac{17\alpha^3}{360^3 \pi^4}  \frac{m_\chi^9}{m_e^8}
	\left[ g_X \left( \frac{ \theta_R^2 }{2} + \frac{m_e^4}{m_E^4} \right)  + \kinmix e  c_W  \right]^2 \, ,
\end{align}
where the large $m_E$ limit was assumed and sub-leading terms of $\cO(\kinmix\, m_e^4/m_E^4)$ and $\cO(\theta_R^2 m_e^2/m_E^2)$ have been neglected. 
The $m_e^4/m_E^4$ term denotes the $E$ loop contribution to the decay amplitude. While the effective field theory treatment of the 3$\gamma$ decay is always valid for $E$ particles running in the loop, the electron contribution in Eq.~\eqref{eq:Gamma3gamma} receives large corrections beyond the effective result for $100\,$keV$\lesssim m_\chi<2m_e$~\cite{McDermott:2017qcg}, which we include in our numerical evaluation of $\Gamma_{\chi\to 3\gamma}$. 
The decay rate into neutrinos is induced from $W$ loops and from tree-level $\chi-Z$ mass mixing for non-vanishing kinetic mixing, which gives (keeping only the dominant logarithmic part at one-loop)
\begin{align}
	\label{eq:Gammanunu}
	\Gamma_{\chi\to \nu\bar\nu}	
\simeq&	\frac{m_\chi}{24\pi}
	\left[ \theta_R^2 \frac{g_Xg^2}{64\pi^2}\frac{m^2_e}{m^2_W} \log\left( \frac{m_E^2 m_W^6}{m^8_e} \right)
	+ \kinmix g_Y \frac{m_\chi^2}{2m_Z^2}
	\right]^2 \, .
\end{align}
The first term in the bracket assumes the large $m_E$ limit, while the second term is the tree-level contribution from mixing with the $Z$ boson. Note that the kinetic mixing does not contribute at loop-level in the $m_\chi\to 0$ limit, since any U(1)$_X$ operator coupling the $\chi_\mu$ field to the LH neutrino current must involve the dark Higgs field. 
For $m_\chi>2m_e$, the dark vector can also decay to $\bar e e$ at tree-level, with a partial width of
\begin{align}
	\Gamma_{\chi\to \bar e e} 
= 	\frac{m_\chi}{24\pi}\sqrt{1-\frac{4m_e^2}{m_\chi^2}}
	\left\{
	\left[ (g_{\chi ee}^L)^2+(g_{\chi ee}^R)^2\right]\left(1-\frac{m_e^2}{m_\chi^2}\right)
	+ 6g_{\chi ee}^L g_{\chi ee}^R \frac{m_e^2}{m_\chi^2}
	\right\}\,.
\end{align}

\begin{figure}[t]
\begin{center}
	\includegraphics[width=0.5\columnwidth]{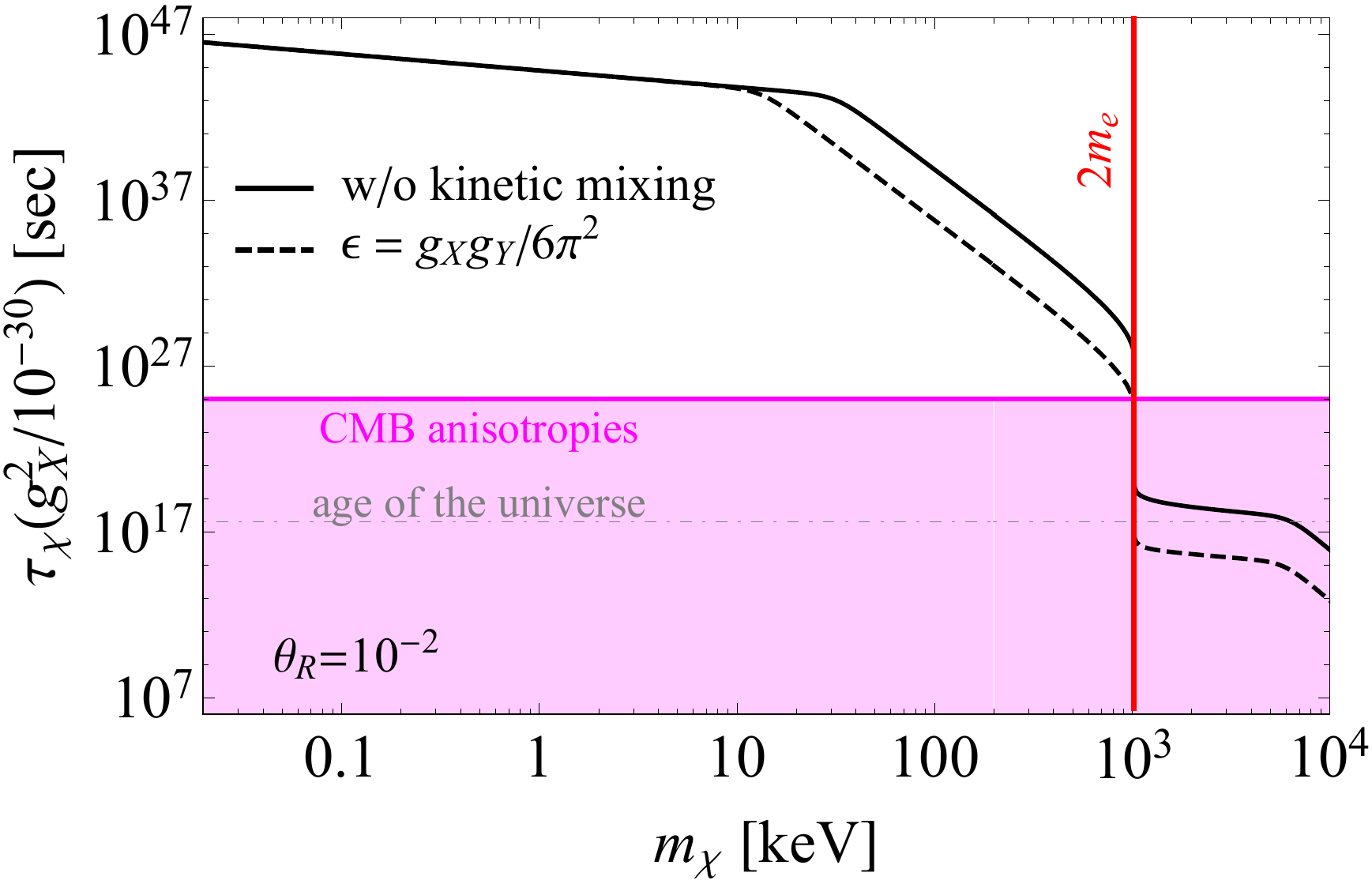}~\includegraphics[width=0.5\columnwidth]{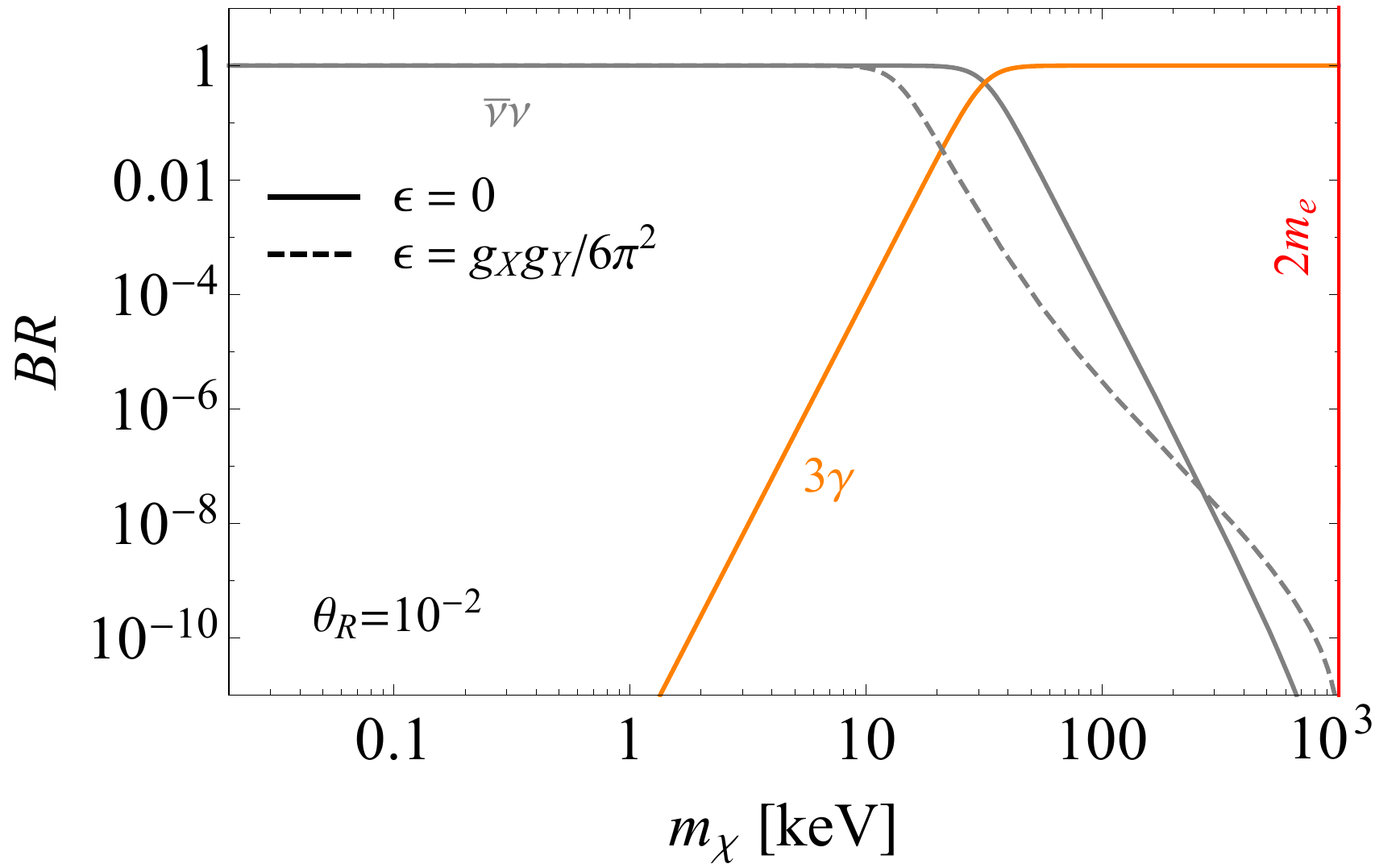}
 	\caption{Lifetime~(left) and BRs~(right) of decaying VDM as function of the VDM mass $m_\chi$, for $m_E=1\,$TeV and $\theta_R=10^{-2}$. 
	Solid\,(dashed) lines assumes zero kinetic mixing\,($\epsilon=g_Xg_Y/(6\pi^2)$). 
	Note that $\tau_\chi \propto g_X^{-2}$ and that all BRs are independent of $g_X$, as well as $\theta_R$ in the absence of kinetic mixing. 
	The dot-dashed line denotes the present age of the Universe and the shaded magenta region is excluded by studies of CMB anisotropies. 
	The vertical red line denotes the electron-positron threshold $m_\chi=2m_e$.}
	\label{fig:VDMdecay}
 \end{center}
\end{figure}

Figure~\ref{fig:VDMdecay} shows the VDM lifetime and branching ratios~(BRs) as function of $m_\chi$. 
For $m_\chi \lesssim 30\,\keV$, the dominant decay channel is into $\nu\bar\nu$, while for larger masses the $3\gamma$ final state dominates up to the electron-positron threshold. 
For $m_\chi >2m_e$, the $e^+e^-$ channel is open and becomes the dominant decay mode, until $m_\chi\approx 25\,$MeV from which the $3\gamma$ final state, whose partial width grows quickly as $m_\chi^9$, dominates again. 
Note that BRs are independent of $g_X$, as well as $\theta_R$ in the absence of kinetic mixing. 
For most of the coupling range that is relevant for reproducing the DM relic density (see below) dark vectors with $m_\chi< 2m_e$ are typically stable on cosmological scale, with a VDM lifetime $\tau_\chi$ that largely exceeds the age of the Universe $\tau_{\rm U} \approx 4.3\times 10^{17}\,$s~\cite{Aghanim:2018eyx}.  Conversely, dark vectors above the electron threshold are typically too short-lived to act as DM, thus we no longer consider this region henceforth.
%
%

%%%%%%%%%%%%%%%%%%%%%%%%%%%%%%%%%%%%%%%%%%%%%%%%%%%%%%%%%%%%%%%%%%%%%%
\section{The VDM relic density }
\label{sec:FreezeIn}
%%%%%%%%%%%%%%%%%%%%%%%%%%%%%%%%%%%%%%%%%%%%%%%%%%%%%%%%%%%%%%%%%%%%%%

This section describes the production of VDM in the early universe. 
We assume that dark sector particles are initially absent from the thermal bath, that is $n_\chi=n_E=0$ at $T=T_R$ where $T_R\ll \Lambda$ is the reheating temperature. 
Because $E$ carries hypercharge, it will be vey quickly develop an equilibrium density from the scattering of hypercharge gauge boson provided $T_R\gg m_E$, which we assume here. Hence, $n_E=n_{\rm eq}$ for $T\lesssim T_R$. 
There are two distinct phases for the production of the $\chi_\mu$ relic density in the early Universe, depending on the size of its couplings to SM fields. If VDM interactions with the thermal bath are too slow, then the relic density will be produced mostly out-of-equilibrium, from a freeze-in mechanism where VDM particles are created by collision of thermal SM (and $E$) particles. Conversely, in the case that such interactions are faster than Hubble, $\chi_\mu$ will reach thermal equilibrium and its relic density will be set by thermal freeze-out.\\

There are two different types of processes that create VDM particles. 
First, $2\to2$ scattering processes with a photon, like $\ell_1\bar \ell_2\to \gamma \chi$ or $\ell_1\gamma\to \ell_2\chi$ with $\ell_i$ being either $e$ or $E$. 
Such processes, creating one VDM particle per collision, are allowed since $\chi_\mu$ is not stable. 
Double production from, for instance, $\ell_1\bar \ell_2 \to \chi\chi$ is also possible. However, it is suppressed by a relative factor of $\sim g_X^2/(4\pi \alpha)$, where $\alpha$ is the fine structure constant, and therefore negligible given the small values of $g_X$ considered. 

Second, VDM can be produced by (in-equilibrium) decay of $E$ particles, $E\to \chi e$. 
This process is parametrically more efficient than $2\to 2$ ones (suppressed by $\alpha$), since it requires one less power of equilibrium density in the initial state, and therefore largely dominates VDM production~\cite{Hall:2009bx}.  
In contrast with VDM scenarios where production is possible through scattering~\cite{Redondo:2008ec}, we show below that the decay channel allows to easily accomodate the observed dark matter abundance without conflicting with constraints from astrophysics and cosmology. \\

In the following, we first consider the regime of freeze-in VDM production from $E$ decays. We then discuss scattering contributions to the freeze-in mechanism, and show that those are neglegible, unless the mixing angle $\theta_R$ which controls the $E$ decay width, is very small (see Fig.~\ref{fig:prodratio}). Finally, we consider the relativistic freeze-out limit. The resulting contour of $\Omega_\chi h^2=0.12$ is shown for illustration in Fig.~\ref{fig:constraints} (black solid line) for $m_E=1\,$TeV and $\theta_R=10^{-2}$.   

%%%%%%%%%%%%%%%%%%%%%%%%%%%%%%%%%%%%%%%%%%%%%%%%%%%%%%%%%%%%%%%%%%%%%%
\subsection{Freeze-in from heavy lepton decay}
%%%%%%%%%%%%%%%%%%%%%%%%%%%%%%%%%%%%%%%%%%%%%%%%%%%%%%%%%%%%%%%%%%%%%%

The Boltzmann equation that determines the VDM number density $n_\chi$ produced from $E\to \chi e$ decay is 
\begin{align}
	\label{eq:Boltzmann}
	\dot n_\chi +3Hn_\chi 
=& 	\int d\Pi_\chi d\Pi_E d\Pi_e\ (2\pi)^4 \delta^{(4)}(p_E-p_\chi-p_e)
	\nonumber \\
&	\times \left[ |\mathcal{M}_{E\to \chi e}|^2 f_E(1-f_e)(1+f_\chi)
	-|\mathcal{M}_{\chi e\to E}|^2 f_\chi f_e(1-f_E)\right]\,,
\end{align}
where $H\approx1.66 \sqrt{g_{\rho}}\, T^2/m_{\rm Pl}$ is the Hubble rate (with $m_{\rm Pl}= 1.22 \times 10^{19}$ GeV the Planck mass and $g_{\rho}$ the number of degrees of freedom related to the energy density $\rho=\pi^2g_\rho T^4/30$), $d\Pi_i = g_i \int d^3p_i/[(2\pi)^32E_i]$ and $f_i$ is the energy distribution of the particle $i$ in the plasma with $g_i$ spin degrees of freedom. 
The first term of the collision integral controls the production of $\chi_\mu$, and the second its depletion from inverse decay. 
The latter is only relevant in a regime of couplings where $\chi_\mu$ reaches equilibrium with the thermal bath. 

In the freeze-in scenario, $f_\chi\simeq 0$ and the inverse decay is negligible. 
Further neglecting quantum statistical effects in the collision integral for the decay $E\to\chi e$, Eq.~\eqref{eq:Boltzmann} becomes
\begin{align}
	\dot n_\chi + 3H n_\chi 
= 	n_E\langle \Gamma_{E\to \chi e}\rangle \, ,
\end{align}  
where $\langle \Gamma_{E\to \chi e}\rangle = \Gamma_{E\to \chi e}m_E \int d^3p_E  (f_E/E_E)/\int d^3p_E f_E$ is the thermal average of the decay rate $\Gamma_{E\to \chi e}$. 
The resulting comoving density $Y_\chi\equiv n_\chi/s$, with $s =2\pi^2 g_{s} T^3/45$ the entropy density expressed in terms the number of degrees of freedom $g_s$, is approximately~\cite{Hall:2009bx} 
\begin{align}
	\label{eq:Boltzmann03}
	Y_{\chi} 
	&\approx 
	 4.3\times 10^{-4} \,\mET^{-2}
	 \left(\frac{\Gamma_{E \to \chi e}}{2.7\times10^{-5}\,\eV}\right)
	  \,,
\end{align}
whose value is chosen such that the VDM relic abundance $\Omega_\chi h^2\approx 2.74\times 10^2Y_\chi\mchik $ accommodates observations, $\Omega_{\rm DM} h^2\approx 0.12$~\cite{Aghanim:2018eyx}, for keV-scale $\chi$; $\mET$ denotes the $E$ mass in units of TeV. In terms of the model parameters we obtain 
\begin{align}
	\label{eq:relicdensity}
	\Omega_{\chi} h^2
&	\approx
	0.12 \frac{\mET}{\mchik}
	\left(\frac{ g_X \, \theta_R}{ 5.3 \times 10^{-17}}\right)^2  \, .
\end{align}
The above derivation assumes that particles in the thermal bath have a Maxwell-Boltzmann energy distribution. One may question the validity of this approximation given that most of the DM is produced at temperatures around $T=m_E/3$ from relativistic $E$ particles in association with an electron whose typical energy is $m_E/2$, comparable to the temperature. When DM is produced from decays of thermal bath particles the effect of quantum statistics is typically important~\cite{Belanger:2018ccd,Heeba:2018wtf}, even more so if DM is produced in association with another thermal bath particle receiving little energy due to phase-space suppression~\cite{Belanger:2018ccd}. Following Ref.~\cite{Belanger:2018ccd}, we find that using Fermi-Dirac statistics for both the heavy lepton and the electron reduces the DM relic density by $\sim 20\%$ relative to a classical treatment. This correction mostly originates from the Pauli-exclusion factor of the electron in Eq.~\eqref{eq:Boltzmann}, which evaluates to $1-f_e\simeq 1-e^{-3/2}\approx 0.8$ at relevant temperatures.

Note that freeze-in production of the observed DM relic is only possible for $m_\chi\gtrsim 60\,$eV. Indeed, $Y_\chi$ cannot exceed the equilibrium value, which implies a lower bound on the VDM mass since $\Omega_\chi h^2\propto m_\chi Y_\chi$. Then, DM is rather produced by a freeze-out mechanism (see below).\\ 

Finally, VDM is also produced from out-of-equilibrium decay of $E$ particles after their density freezes out. This late-time contribution to the density of $\chi$ particles is given by the density of $E$ (and $\bar E$) particles at freeze-out $Y_E^{\rm fo}$, multiplied by the $E\to \chi e $ branching ratio. 
$E$ and $\bar E$  are kept in chemical equilibrium with the thermal bath by hypercharge interactions and their freeze-out density is set by $E\bar E\to BB$ annihilation. 
The annihilation is $s$-wave with a thermally-averaged cross section of $\langle\sigma_{E\bar E\to BB}v\rangle\simeq g_Y^4/(8\pi m_E^2)$. 
A standard freeze-out calculation~\cite{Gondolo:1990dk} then yields $Y_E^{\rm fo}\simeq Y_E^{\rm eq}(T\approx m_E/26)\approx 3\times 10^{-12}$, with little dependence on the value of $m_E$.\footnote{For $m_E$ close to the TeV scale, freeze-out typically occurs in the broken SU(2)$_L\times$U(1)$_Y$ phase where $B$ is no longer a proper mass eigenstate. However, $E$ has vector-like gauge interactions and the finiteness of the $Z$ boson mass only induces a small correction of $\sim m_Z^2/m_E^2\ll 1$ to the total annihilation cross section.} 
In view of Eq.~\eqref{eq:Boltzmann03}, the fraction of VDM produced from out-of-equilibrium decay is therefore totally negligible relative to the freeze-in contribution.   

%%%%%%%%%%%%%%%%%%%%%%%%%%%%%%%%%%%%%%%%%%%%%%%%%%%%%%%%%%%%%%%%%%%%%%  
\subsection{Subleading scattering contributions}
%%%%%%%%%%%%%%%%%%%%%%%%%%%%%%%%%%%%%%%%%%%%%%%%%%%%%%%%%%%%%%%%%%%%%%

The discussion above is based on $\chi$ production from $E$ decay, and neglects sub-leading contributions from scattering. 
We provide here a more detailled analysis of the latter in order to establish the robustness of this approximation. 
While we use the full expressions for the scattering amplitudes in our numerical calculations, the corresponding DM yields can be estimated as follows. 
In the relativistic limit the $2\to 2$ scattering rate scales as $n^{\rm eq}\langle \sigma v\rangle \sim \lambda^2 \, T $ where $\lambda$ generically denotes the coupling constant of $\chi$ (including possible enhancement factor due to the longitudinal polarization of $\chi$) with the thermal bath and $n^{\rm eq}\sim T^3$ is the equilibrium number density of the initial states. 
This reaction rate is to be compared with the Hubble rate which, during the radiation-dominated era, scales like $H\sim T^2/m_{\rm Pl}$. 
Hence, in this regime, DM production is more efficient the lower the temperature. 
Moreover, as soon as $T$ decreases below $m_{\rm h}$, the mass of the heaviest particle involved in the scattering, the reaction rate becomes exponentially suppressed and DM production quickly stops. 
Consequently, the final DM yield is approximately given by 
\begin{align}
	Y_\chi^{2\to 2} 
	\sim 
	\frac{\lambda^2}{\bar g^{3/2}(m_{\rm h})} \frac{m_{\rm Pl}}{m_{\rm h}}\,,
\end{align}
where $\bar g^{3/2}\equiv g_s^2/g_*^{1/2}$ with $g_*^{1/2} = (g_s/g_\rho^{1/2})[1+1/3(d\log g_s/d\log T)]$. 
Similarly, the decay contribution is approximately $Y_\chi^{\rm decay}\sim \kappa^2 m_{\rm Pl}/[m_E\,\bar g^{3/2}(m_E)]$ where $\kappa$ is a generic decay constant ($\Gamma\propto \kappa^2m_E$). 

There are three types of $2\to 2$ processes which involve a photon and either two $E$ particles, two electrons or one $E$ particle and an electron. We henceforth denote the total yield associated with these three contributions as $Y_\chi^{EE}$, $Y_\chi^{ee}$ and $Y_\chi^{Ee}$, respectively. 

In the absence of kinetic mixing, the effective couplings are $\lambda_{EE}\sim \sqrt{4\pi\alpha} g_{\chi EE}^V$, $\lambda_{ee}\sim\sqrt{4\pi\alpha} g_{\chi ee}^A (m_e/m_\chi)$ and $\lambda_{Ee}\sim\sqrt{4\pi\alpha} g_{\chi Ee}^R (m_E/m_\chi)$. 
The additional $m_\ell/m_\chi\gg 1$ factor in processes involving electrons stems from the fact the latter only couple to longitudinaly-polarized $\chi$, which corresponds to the eaten Goldstone boson of the spontaneously broken U(1)$_X$ gauge group, through the Yukawa operator in $\cL_{\rm portal}$. Note that only the axial part of $g_{\chi ee}$ is coupled to the longitudinal polatization, while the vector part is not enhanced by $m_e/m_\chi$. This is so because, in the vector-like limit, the global U(1)$_X$ symmetry is preserved by the gauge boson mass term. Thus, the associated current is still conserved and the $1/m_\chi^2$ term from the polarization sum does not contribute to the amplitude squared. 
Similarly, processes with two $E$ particles are controlled by U(1)$_X$ gauge interactions, thus they are not enhanced by the longitudinal polarization factor. 
Furthermore, processes with at least one $E$ particle shut off rather early, near $T\sim m_E$, due to the heaviness of $E$, while processes with two electrons perdure until $T\sim m_e\ll m_E$ and are thus enhanced by a relative factor of 
$m_E/m_e\times \bar g^{3/2}(m_E)/\bar g^{3/2}(m_e)\sim 10^8\times (m_E/{\rm TeV})$,  using  $\bar{g}^{3/2}(m_{E}) / \bar{g}^{3/2}(m_{e}) \approx 40$. 
Hence, 
the yields from the different scattering processes are expected to scale as
\begin{align}
	\label{eq:Y22ratioseps0}
	Y_\chi^{Ee} 
	\sim 
	\frac{4\pi\alpha g_X^2\theta_R^2}{\bar g^{3/2}(m_E)}\frac{m_{\rm Pl}m_E}{m_\chi^2 }
	\sim 
	\theta_R^2 \frac{ m_E^2}{m_\chi^2}Y_\chi^{EE}
	\sim 
	\frac{4}{\theta_R^2}
	\frac{\bar g^{3/2}(m_e)}{\bar g^{3/2}(m_E)}\frac{m_E}{m_e}Y_\chi^{ee}\,.
\end{align}
Taking $m_E=1\,\TeV$,  $Y_\chi^{ee}$ is thus always subdominant to $Y_\chi^{Ee}$, as well as to $Y_\chi^{EE}$ whenever $\theta_R\lesssim 10^{-4}(\mchik)^{1/2}$, and $Y_\chi^{Ee}$ dominates over $Y_\chi^{EE}$ for $\theta_R\gtrsim 10^{-9}\mchik$. In turn, the ratio $R$ of scattering to decay contributions to DM production
\begin{align}
	R
	\equiv 
	\frac{Y_\chi^{2\to 2}}{Y_\chi^{\rm decay}}\,,
\end{align}
is (neglecting the subleading contribution from two-electron processes)
\begin{align}
	\label{eq:Rest}
	R (\kinmix=0) 
&	\sim 
	4\pi\alpha\left( 1+ \frac{m^2_\chi}{\theta_R^2 m^2_E } \right)  \, ,
\end{align}
where the effective coupling constant for decay $\kappa \sim g_{\chi Ee}^R (m_E/m_\chi)$ is also enhanced because of the longitudinal polarization. This shows that DM production is dominated by decay unless $\theta_R\lesssim 3\times 10^{-10}\mchik$, in which case DM is mostly produced by scattering processes of $E$ particles only.\\

Including a nonzero kinetic mixing  $\kinmix=g_Xg_Y/(6\pi^2)$, the only significant change is in the yield from two-electron processes $Y_\chi^{ee}$. The $\chi$ coupling to electron pairs has now a vector-like part whose 
contribution to the scattering cross section is dominated by transverse polarizations, as argued above. Consequently, the effective coupling is parametrically $\lambda_{ee}\sim \sqrt{4\pi\alpha}\sqrt{(g_{\chi ee}^V)^2+(g_{\chi ee}^A)^2(m_e/m_\chi)^2}$ and the scaling in Eq.~\eqref{eq:Y22ratioseps0} is modified as
\begin{align}
	Y_\chi^{ee}(\kinmix \neq 0) 
	\sim 
	\left(1
	+\frac{16\alpha^2}{9\pi^2\theta_R^4}\frac{m_\chi^2}{m_e^2}
	\right)Y_\chi^{ee}(\kinmix=0)\,,
\end{align}
where subleading term of $\cO(\theta_R^2)$ has been neglected in $g^V_{\chi ee}$ since typically $\theta_R\ll \sqrt{4\alpha/(3\pi)}\approx 0.06$. Now, taking $m_E=1\,\TeV$, $Y_\chi^{ee}$ always dominates over $Y_\chi^{EE}$, as well as $Y_\chi^{Ee}$ whenever $\theta_R\lesssim 3\times 10^{-8}\mchik$ and the scattering-to-decay ratio becomes, approximately,
\begin{align}\label{Rratiokinmix}
	 R (\kinmix\neq 0) 
&	\sim 
	4\pi\alpha\left(1+\frac{16\alpha^2}{9\pi^2 \theta_R^2} \frac{\bar g^{3/2}(m_E)}{\bar g^{3/2}(m_e)} \frac{m_\chi^2}{m_em_E}\right)\,.
\end{align}
Therefore, decay dominates DM production unless $\theta_R\lesssim 10^{-8}\mchik$, in which case the DM relic is set by scattering processes with two electrons.\\
\begin{figure}[t]
\begin{center}
	 \includegraphics[width=0.6\columnwidth]{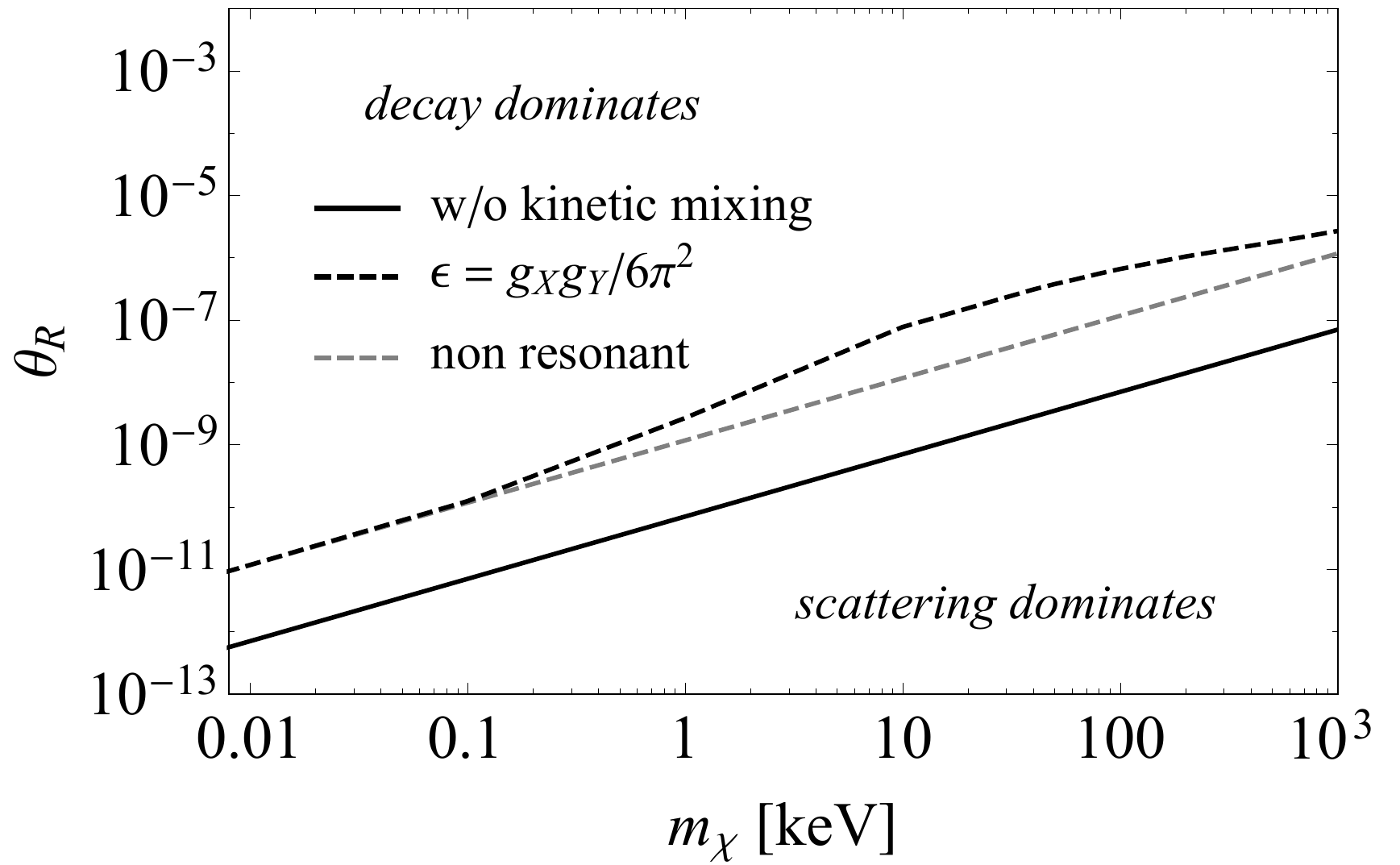}
 	\caption{Regions where VDM production is dominated by decay ($R \equiv Y_\chi^{2\to 2}/Y_\chi^{\rm decay}<1)$ and $2\to 2$ 
	scattering processes ($R>1$) in the $m_\chi-\theta_R$ plane, for cases without kinetic mixing~(solid) and with 
	$\kinmix=g_Xg_Y/(6\pi^2)$~(dashed), and assuming $m_E =1\,\TeV$. The  gray curve shows the scattering-to-decay ratio without the resonant enhancement from mixing with the photon in the plasma.}
	\label{fig:prodratio}
 \end{center}
\end{figure}

The above estimates reproduce, up to $\cO(1)$ factors, the full numerical calculation of the $R$ ratio whose results are presented in Fig.~\ref{fig:prodratio} for $\kinmix=0$ and $\kinmix=g_X g_Y/(6\pi^2)$.\\

So far we have assumed that yields from $2\to 2$ processes are dominated by temperatures comparable to the heaviest particle mass involved in the scattering. While this is generically true in vacuum, it is possible to get larger resonant contributions at higher temperatures once plasma effects are taken into account~\cite{Redondo:2008ec}. Photons in the plasma acquire a nonzero  mass of $m_\gamma(T)=2\alpha T^2/3$ from interacting with thermal electrons~\cite{Weldon:1982aq}. As a result, the scattering yield is resonantlty enhanced, by mixing with the photon when its thermal mass equals that of the VDM. The VDM yield from such resonant photon conversion effect is particularly important for processes with two electrons when $m_\chi<2m_e$~\cite{Redondo:2008ec}. Note that resonant photon conversion is negligible for processes involving the heavier $E$ particles since the temperature $T_r$ at which resonance occurs is $\sim 8m_\chi\ll m_E$ typically. %in which case the resonance contribution to the $Y_{\chi}^{ee}$ yield is approximately given by
%\begin{align}
%Y_\chi^{ee}\big_{\rm res}\simeq \frac{\lambda_{ee}^2}{\bar g^{3/2}(T_r)}
%\end{align}
%\begin{align}
%Y_\chi^{ee}\big|_{\rm res}\sim \frac{\lambda_{ee}^2}{\bar g^{3/2}(T_r)}\frac{m_{\rm Pl}}{T_r}F_r\,,
%\end{align}
%where $F_r\sim 2\pi^2/[3\alpha\log(4T_r^2/m_e^2)]$~\cite{Redondo:2008aa}. 
Since it originates from mixing with the photon only the vector part of the electron coupling is involved. In the absence of kinetic mixing, the two-electron process contribution to the scattering yield is  largely subdominant and its resonant enhancement is not relevant. It is however relevant in the case of nonzero kinetic mixing where $Y_\chi^{ee}$ dominates over $Y_\chi^{EE}$ and $Y_\chi^{Ee}$. We calculated the resonant photon conversion contribution following Ref.~\cite{Redondo:2008ec} and added it to the non resonant one evaluated without plasma effects. As shown in  Fig.~\ref{fig:prodratio} the yield from photon conversion in the plasma typically exceeds the non resonant contribution by one to two orders of magnitude depending on the VDM mass. It is not significant for $m_\chi\lesssim 1\,$keV, which corresponds to $T_r\lesssim m_e/3$, due to the Boltzmann-suppressed electron density in the plasma at these temperatures.

%%%%%%%%%%%%%%%%%%%%%%%%%%%%%%%%%%%%%%%%%%%%%%%%%%%%%%%%%%%%%%%%%%%%%%
\subsection{The relativistic freeze-out limit}
\label{sec:FreezeOut}
%%%%%%%%%%%%%%%%%%%%%%%%%%%%%%%%%%%%%%%%%%%%%%%%%%%%%%%%%%%%%%%%%%%%%%

For large enough values of $\Gamma_{E\to \chi e}$ 
the production of $\chi_\mu$ can be so efficient that the contribution from the inverse decay $\chi e\to E$ becomes relevant, eventually bringring VDM in equilibrium with the thermal bath if $\langle \Gamma_{E\to \chi e} \rangle\gg H$. In this case, the VDM abundance is rather set by a freeze-out mechanism. 

For this to occur the rate of VDM interactions with the thermal bath must be faster than the Hubble rate before $E$ becomes non-relativistic, which corresponds to $n_E/n_\chi\times \langle \Gamma_{E\to e\chi}\rangle/H \gtrsim1$ at $T\approx m_E/3$. Given the partial decay width of $E\to e\chi$ in Eq.~\eqref{E2eXwidth},
 the above thermalization condition implies a lower bound on the Yukawa portal coupling of~\footnote{The reaction rate is understood here as being the sum of $E$ and $\bar E$ decays, namely $n_E\equiv g_E\int d^3p_E f_E$ with $g_E=4$.}
\begin{align}
	x_E\gtrsim 3.6\times 10^{-7} \sqrt{\mET}\,,
\end{align}
or, equivalently,
\begin{align}
	g_X 
\gtrsim 	 2.5\times10^{-14} \frac{\mchik}{ 
	\sqrt{\mET}}
	\left( \frac{10^{-2}}{\theta_R} \right) \, .
\end{align}

In this regime, the Boltzmann equation receives a contribution from the inverse decay process and reads~\footnote{We assume that there is an efficient energy transfer between the SM and $\chi_\mu$ in order to write the reverse process contribution in terms of $n_\chi/n_\chi^{\rm eq}$ in Eq.~\eqref{eq:Boltzmann04}. While this is not generically the case, it does however hold when $\chi_\mu$ is close to thermal equilibrium, which is the limit of interest here.}
\begin{align}
	\label{eq:Boltzmann04}
	\dot n_\chi + 3H n_\chi 
= 	n_E\langle \Gamma_{E\to \chi e}\rangle 
	\left(1-\frac{n_\chi}{n_\chi^{\rm eq}}\right)\,,
\end{align} 
where $n_\chi^{\rm eq}$ is the VDM equilibrium density. Chemical equilibrium with the SM is maintained as long as $E$ is abundant in the thermal bath, for $T\gg m_E$. When $T\lesssim m_E$, $E$ is no longer produced efficiently by thermal collisions and its density becomes exponentially suppressed. Eventually, this triggers the decoupling of  $\chi_\mu$, while still relativistic, when the decay rate becomes slower than Hubble at $T=T_f\equiv m_E/x_f$ with $x_f\gtrsim 3$.  
Then, 
$Y_\chi\approx Y_\chi^{\rm eq}(T_f\gg m_\chi) $
 is independent of the VDM mass and only mildly sensitive to the portal coupling $x_E$, through the value of $x_f$. This is in sharp contrast to the canonical non-relativistic freeze-out scenario where the DM equilibrium density falls exponentially before decoupling, thus inducing a strong dependence on the DM coupling to the thermal bath.     
 
Relativistic freeze-out occurs approximately when $n_E/n_\chi^{\rm eq}\langle \Gamma_{E\to \chi e}\rangle/H\lesssim 1$. Approximating the $E$ density as $n_E\simeq g_Ee^{-m_E/T}[m_E/(2\pi T)]^{3/2}$ and $\langle \Gamma_{E\to \chi e}\rangle\simeq \Gamma_{E\to \chi e}$, valid for $T\lesssim m_E$, yields
\begin{align}\label{eq:xf}
	x_f  \simeq \log\left(\frac{4.7g_E\Gamma_{E\to \chi e} m_{\rm Pl}g_*^{1/2}}{g_\chi m_E^2g_s}\right)+\frac{7}{2}\log\left[\log\left(\frac{4.7g_E\Gamma_{E\to \chi e} m_{\rm Pl}g_*^{1/2}}{g_\chi m_E^2g_s}\right)\right]\,,
\end{align}
where $g_E=4$ and $g_\chi=3$ are spin degrees of freedom of $E$ (and $\bar E$) and $\chi$ particles. The variation with temperature of the numbers of degrees of freedom is typically negligible compared to the exponential falling of the $E$ number density,  and $g_s$ and $g_*^{1/2}$ are assumed constant and evatuated at $T_f$ in Eq.~\eqref{eq:xf}.  The larger the decay width, the later VDM freezes out, as expected. For a leptonic portal coupling in the range $3.6\times 10^{-7}\sqrt{\mET}<x_E<0.1$, one finds approximately $3.2\lesssim x_f\lesssim 38-\log\mET$. 

Consequently, $\Omega_\chi$ matches the observed abundance of DM only for a very narrow range of VDM masses, whose width reflects the mild sensitivity of $Y_\chi^{\rm eq}$ to the value of $x_f$ (that is to the value of the VDM-to-SM coupling). Since $\chi$ is relativistic at freeze-out, $n_\chi^{\rm eq}/T^3$ is constant and this dependence arises merely from the (possible) change in the number of entropy degrees of freedom during freeze-out. Assuming $\Omega_\chi h^2\approx 0.12$ yields 
\begin{align}\label{eq:massFO}
	m_\chi \approx 67 \, \eV \left[\frac{g_s(m_E/x_f)}{106.5}\right]\,,
\end{align}
where we use the value $g_s\approx 106.5$ for $m_E=1\,$TeV and $x_f=3.2$.  
For $m_\chi$ above the value in Eq.~\eqref{eq:massFO}, VDM is overabundant, while for lighter masses $\chi$ cannot explain all the observed DM within the freeze-out regime.

Note that for relatively low $E$ mass values, a larger value of $x_f$ could imply a slightly different  number of degrees of freedom . Consider for instance the extreme case of $x_E=0.1$, yielding a late freeze-out at $T_f\approx m_E/38\approx 26\,$GeV for $m_E=1\,$TeV. At this temperature, the top quark, the Higgs and the weak gauge bosons have decoupled from the thermal bath and $g_s(T_f)\approx 89.6$. Then,  a lower mass of $m_\chi\approx 56\,$eV is required to accomodate the DM relic density. Conversely, for $m_E\gtrsim 6.3\,$TeV, freeze-out always occurs while all SM states are active and $g_s$ does not change. Hence, in this case, the relic density predicts a single VDM mass scale of $m_\chi\approx 67\,$eV, independently of VDM couplings.\\

At temperatures $T\ll m_E$, VDM could still be kept in thermal equilibrium through, now dominant, $2\to 2$ scattering processes such as $\bar ee\to\chi\gamma$ and $\gamma e\to\chi e$ (and their inverse). 
However, due to the heaviness of $E$ and the smallness of the $\chi$ coupling to electrons, these two processes are found very inefficient and too slow 
to keep VDM in thermal equilibrium with the SM bath after the decoupling of $E$ particles.

%%%%%%%%%%%%%%%%%%%%%%%%%%%%%%%%%%%%%%%%%%%%%%%%%%%%%%%%%%%%%%%%%%%%%%
\section{DM phenomenology and collider signatures}
\label{sec:signatures}
%%%%%%%%%%%%%%%%%%%%%%%%%%%%%%%%%%%%%%%%%%%%%%%%%%%%%%%%%%%%%%%%%%%%%%

Our VDM model has several possible experimental signatures. 
Those include DM absorption in direct detection searches, indirect astrophysical and cosmological probes from stellar cooling, $\chi\to3\gamma$ decays and cosmic microwave background~(CMB). All of them are mostly sensitive to $g_{\chi ee}$. In addition, we consider also possible constraints from big-bang nucleosynthesis~(BBN) and structure formation.

The VDM relic abundance fully determines the coupling $ g_{\chi Ee} \sim g_X \theta_R$ for fixed $m_\chi$ and $m_E$, see Eq.~\eqref{eq:relicdensity}. 
In absence of kinetic mixing, the VDM coupling to electrons scales as $g_{\chi ee}^R \sim g_X \theta_R^2$, thus all VDM experimental signatures are parametrically suppressed. Therefore, VDM can easily saturate the observed relic density while evading constraints from existing searches.  
However, in the presence of a nonvanishing kinetic mixing, the direct correlation between the relic density and experimental signatures is altered, since now $\epsilon$ contributes to $g_{\chi ee}$, but not to $g_{\chi Ee}$. 
Assuming $\kinmix=g_Xg_Y/(6\pi^2)$, $g_{\chi ee}$ is dominated by the kinetic mixing for $\theta_R < \sqrt{2\alpha /(3\pi)}\approx 4\times 10^{-2}$.
Therefore, for smaller  $\theta_R$ values, saturation of the relic density requires larger values of $g_X$, resulting in an enhancement of the VDM signals in experiments. 

\begin{figure}[t]
\begin{center}
  	\includegraphics[width=0.5\columnwidth]{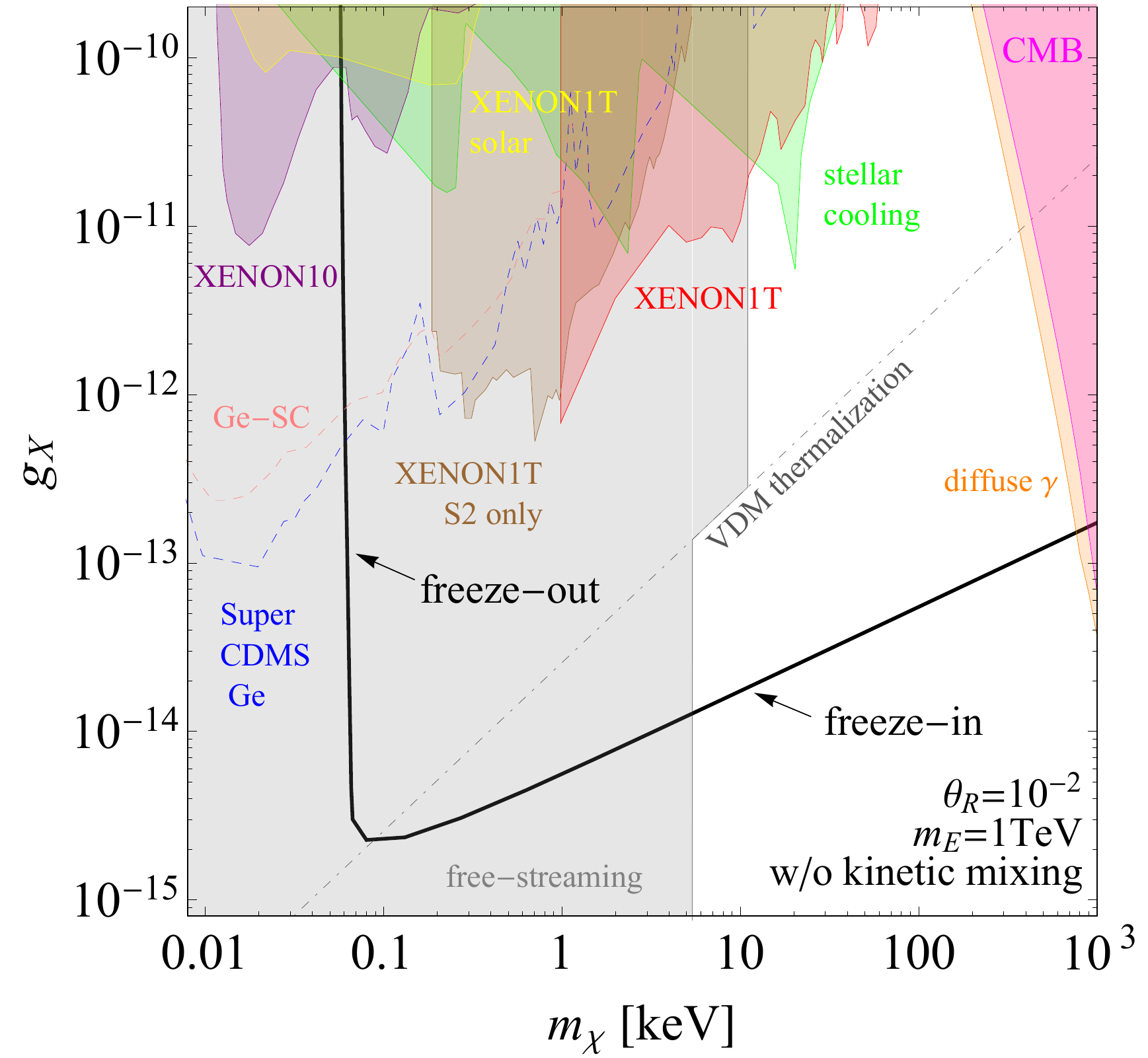}~
    	\includegraphics[width=0.5\columnwidth]{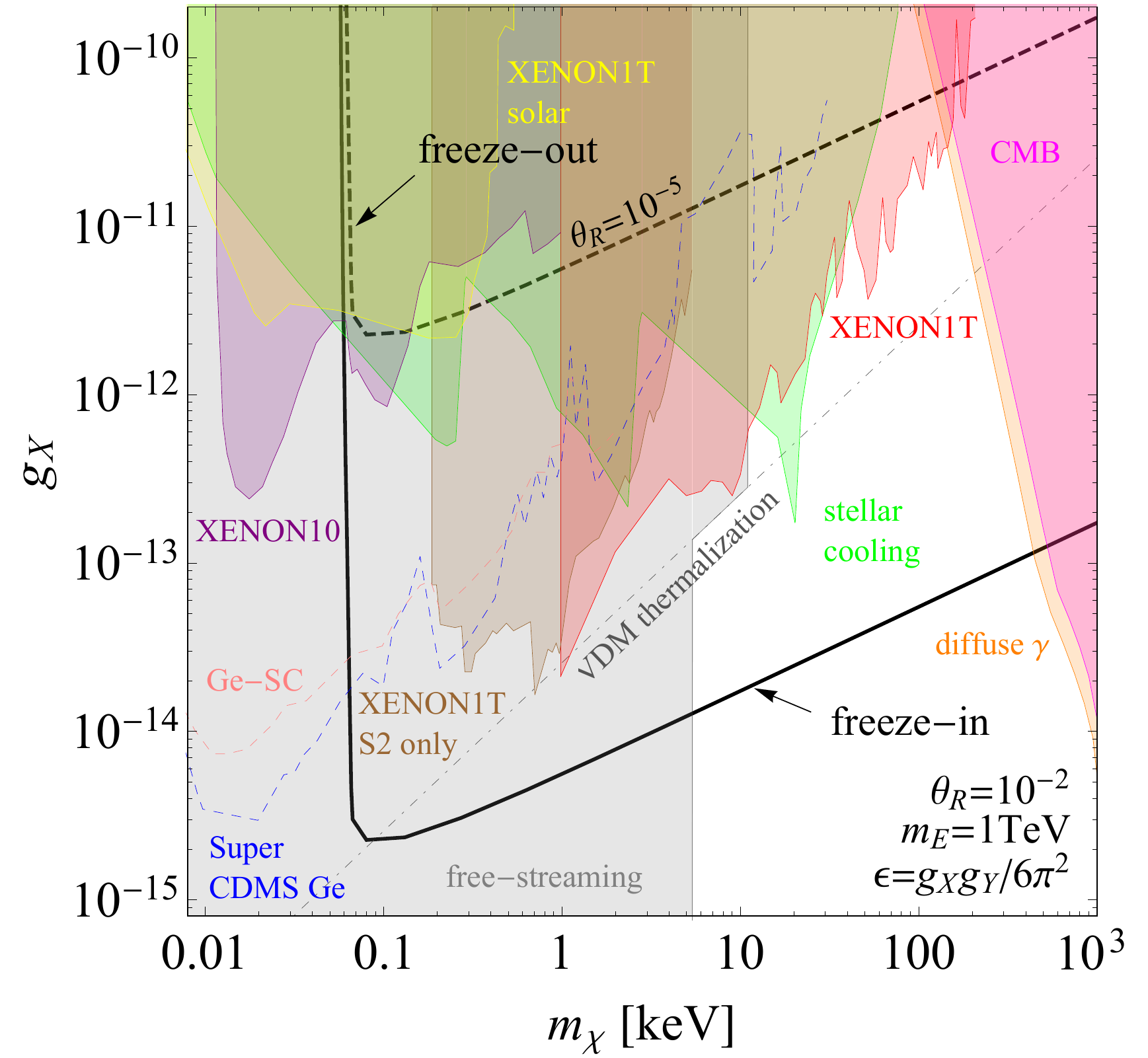}
  	\caption{
	Constraints on VDM, $\chi_\mu$, in the $m_\chi-g_X$ plane for $m_E=1\,\TeV\,$, $\theta_R=10^{-2}$ and $\epsilon=0$\,(left) or $\epsilon=g_Xg_Y/6\pi^2$\,(right). 
	The solid black line denotes the $\Omega_\chi h^2=0.12$ contour above\,(below) which VDM is over\,(under) abundant. 
	Above the dot-dashed line VDM is in chemical equilibrium with the thermal bath and produced by freeze-out while relativistic; 
	below, VDM is produced by freeze-in through $E\to \chi e$ decay. 
	Shaded areas show regions excluded by searches of VDM absorption in direct detection experiments, including recast~\cite{An:2014twa} from XENON10~\cite{Angle:2011th}~(purple) and XENON1T with electronic recoil data~\cite{Aprile:2020tmw}~(red) or ionisation signals~(S2) only~\cite{Aprile:2019xxb}~(brown), solar emission (yellow), stellar cooling constraints~\cite{An:2014twa}, CMB anisotropies~\cite{Poulin:2016anj}~(magenta), diffuse $\gamma$-ray background observations~\cite{Yuksel:2007dr} following Ref.~\cite{An:2014twa}~(orange) and structure formation~(gray). 
	Dashed lines denote projected sensitivities from future direct detection experiments like SuperCDMS, with Ge target~\cite{Bloch:2016sjj}~(blue), and using Ge semiconductor target~\cite{Hochberg:2016sqx}~(pink).
}
  	\label{fig:constraints}
\end{center}
\end{figure}

We review below the searches relevant to our model and discuss the expected VDM signals in more details. The current constraints and future prospects on the VDM model are summarized in Fig.~\ref{fig:constraints} in the $m_\chi - g_X$ plane for $m_E=1\,\TeV$, $\theta_R=10^{-2}$ and assuming $\epsilon=0$ or $\epsilon=g_Xg_Y/6\pi^2$.

%%%%%%%%%%%%%%%%%%%%%%%%%%%%%%%%%%%%%%%%%%%%%%%%%%%%%%%%%%%%%%%%%%%%%%
%%%%%%%%%%%%%%%%%%%%%%%%%%%%%%%%%%%%%%%%%%%%%%%%%%%%%%%%%%%%%%%%%%%%%%

%%%%%%%%%%%%%%%%%%%%%%%%%%%%%%%%%%%%%%%%%%%%%%%%%%%%%%%%%%%%%%%%%%%%%%
\subsection{Direct detection}
\label{sec:DD}
%%%%%%%%%%%%%%%%%%%%%%%%%%%%%%%%%%%%%%%%%%%%%%%%%%%%%%%%%%%%%%%%%%%%%%

Relic DM particles at the MeV scale or below are detectable through their interactions with atomic bound electrons. DM could either scatter or even be absorbed. In  models where DM couples only feebly with the SM, scattering is typically very suppressed relative to absorption as the cross section is quartic in small couplings and DM only deposites a fraction of its kinetic energy $m_\chi v_\chi^2/2$ where $v_\chi\sim 10^{-3}$ is the average DM velocity in our galaxy. The VDM-electron scattering cross section~\cite{Essig:2011nj} is typically extremely small\footnote{In the presence of kinetic mixing, the scattering amplitude receives extra contributions coming from the vector part of the electron coupling which is suppressed by a factor of $\cO(m_\chi^2/m_e)$ due U(1)$_X$ current conservation. The amplitude from virtual $E$  exchange is relatively suppressed by a factor of $(g_{\chi Ee}^R/g_{\chi ee}^R)^2m_\chi m_e/m_E^2\approx 5\times 10^{-16}\theta_R^{-2}\mchik^2\mET^{-4}$ and negligible.} $\sigma_e\simeq (g_{\chi ee}^R)^4/(8\pi m_\chi^2)\sim \cO(10^{-90}\,$cm$^2)$ for $\theta_R=10^{-2}$ and too many orders of magnitude below experimental sensitivities~\cite{Aprile:2019xxb,Barak:2020fql} to be observed. 

Instead, absorption of VDM particles is more promissing. Not only DM deposits its entire mass energy when absorbed by a material but the associated cross section  scales only quadratically in small couplings. The absorption signal is typically dominated by the vector part of the DM-electron coupling, contributions from the vector-axial part being suppressed by powers of the DM momentum-to-mass ratio. 
Here we adapt several existing bounds on absorption of canonical dark photon DM, namely scenarios where the dark photon couples to the SM only through kinetic mixing with the photon field strength, using the replacement rule $e\kinmix\to g_{\chi ee}^V\simeq g_X\theta_R^2/2+ec_W\kinmix$. 

Of relevance are dark photon bounds extracted from recasts~\cite{An:2014twa,Bloch:2016sjj} of XENON10~\cite{Angle:2011th} and XENON100~\cite{Aprile:2014eoa} data, from XENON1T using electronic recoil data~\cite{Aprile:2020tmw} or ionisation signals~(S2) only~\cite{Aprile:2019xxb}, and from DAMIC~\cite{Aguilar-Arevalo:2016zop} and SENSEI~\cite{Tiffenberg:2017aac,Abramoff:2019dfb,Barak:2020fql} experiments.
XENON10, XENON100 and XENON1T are sensitive to absorption of DM whose mass is above their photoelectric threshold, namely $m_\chi \ge E_{\rm th} = 12.13\,\eV$, while DAMIC and SENSEI have sensitivity to lower masses less relevant for the VDM parameter space under consideration.  
Direct detection experiments also serve as a target for dark photon emission from the Sun~\cite{An:2013yua}. We use here the latest bound from a recast~\cite{An:2020bxd} of the XENON1T exclusion on solar emission~\cite{Aprile:2020tmw}. Note that bremstrahlung contribution from the dark Higgs $\varphi$ is negligible here due to its heavy mass. 
 
Resulting constraints and projections for our VDM scenario  are presented in Fig.~\ref{fig:constraints} on the $m_\chi-g_X$ plane for fixed $\theta_R=10^{-2}$. The shaded regions are excluded by DM searches by XENON10 (purple) and XENON1T (brown and red) and solar emission of new bosons by XENON1T (yellow). For sake of illustration of near future sensitivities on DM absorption in direct detection experiments, we show projections (dashed lines) of Ge semiconductor target with 1\,kg-year exposure~\cite{Hochberg:2016sqx} (pink) and SuperCDMS Ge with 20\,kg-years exposure~\cite{Bloch:2016sjj} (blue).

%%%%%%%%%%%%%%%%%%%%%%%%%%%%%%%%%%%%%%%%%%%%%%%%%%% %%%%%%%%%%%%%%%%%%%
\subsection{Indirect detection}
\label{sec:ID}
%%%%%%%%%%%%%%%%%%%%%%%%%%%%%%%%%%%%%%%%%%%%%%%%%%%%%%%%%%%%%%%%%%%%%%

Late $\chi\to 3 \gamma$ decays would contribute to the diffuse $\gamma$-ray background~\cite{Redondo:2008ec,An:2014twa}.
Here, we follow Ref.~\cite{An:2014twa} and conservatively require that the sum of galactic and cosmological (extragalactic) VDM contributions to the diffuse $\gamma$-ray flux never exceeds the observed one~\cite{Yuksel:2007dr}.
For model parameters such that VDM production overshoots observations, we nevertheless assume that the VDM density at late time is $\Omega_\chi h^2=0.12$. In other words, we implicitly assume that the VDM density is subsequently diluted to the observed value by some unspecified mechanism. Conversely, for parameters yielding a VDM density below observations, we scale down the expected signal of Ref.~\cite{An:2014twa} by a factor of $\Omega_\chi h^2/0.12$. The resulting exclusion is shown in Fig.~\ref{fig:constraints} as an orange-shaded region. We note that indirect detection is a rather weak constraint on the region of parameter space consistent with the relic density, excluding only a small region below the electron-positron threshold, $m_\chi\lesssim 2m_e$. 

%%%%%%%%%%%%%%%%%%%%%%%%%%%%%%%%%%%%%%%%%%%%%%%%%%%%%%%%%%%%%%%%%%%%%%
\subsection{Astrophysical and cosmological constraints}
%%%%%%%%%%%%%%%%%%%%%%%%%%%%%%%%%%%%%%%%%%%%%%%%%%%%%%%%%%%%%%%%%%%%%%

We consider here possible constraints arising from VDM altering BBN and CMB physics, LSS formation and stellar evolution.

\paragraph{BBN}
In constrast with usual WIMP scenarios, where DM is produced cold long before the onset of BBN at $T\sim\cO($MeV$)$, here VDM is produced hot and remains relativistic until after the formation of primordial nuclei. 
As a result, VDM would provide an additional contribution $\Delta N_{\rm eff}^\chi$ to the effective number of neutrinos, $N_{\rm eff}$. 
A too large $N_{\rm eff}$ would hasten Hubble expansion compared to standard cosmology and thus modify light element abundances~\cite{Shvartsman:1969mm, Steigman:1977kc}. 
  
Since VDM is not in equilibrium with the thermal bath at the time of BBN, its contribution to the effective number of neutrinos is~\cite{Sarkar:1995dd}
\begin{align} 
	\label{eq:effectiev_number}
	\Delta N_{\rm eff}^\chi
=	\frac{120}{7\pi^2}\frac{\rho_\chi}{T_\nu^4} 
=	 \frac{4}{7} g_\chi\left( \frac{T_\chi}{T_\nu} \right)^4 \, ,
\end{align}
where $\rho_\chi$ is the VDM energy density, $T_\chi$ is the VDM temperature (in the case of kinetic equilibrium) and $T_\nu\sim \MeV$ is the temperature of neutrino decoupling. 
In the case that VDM ever was in thermal equilibrium with the SM, as in the freeze-out regime, its temperature after decoupling is fixed by entropic considerations. Once VDM decouples at $T=T_f=m_E/x_f$, its entropy and that of the SM are separately conserved. It then follows that the ratio $T_\chi/T$ is tied to the variation of the number of entropy degrees of freedom of the SM between $T$ and $T_f$ as~\cite{Olive:1980wz,Sarkar:1995dd}
\begin{align}
	\label{eq:Tratio}
	\frac{T_\chi}{T} 
= 	\left[\frac{g_s(T)}{g_s(T_f)}\right]^{1/3}\,.
\end{align}
Taking $m_E=1\,$TeV, the lowest decoupling temperature is $T_f\approx 26\,$GeV (corresponding to the case of large portal coupling, $x_E=0.1$). Then, combining Eqs.~\eqref{eq:Tratio} and~\eqref{eq:effectiev_number}, the VDM contribution to the effective number of neutrinos does not exceed 
\begin{align}
	\label{eq:NeffFO}
	\Delta N_{\rm eff}^\chi \lesssim 0.1 \, .
\end{align}  
Larger $E$ masses imply an earlier VDM decoupling, and a slightly smaller contribution to $N_{\rm eff}$. 

In the freeze-in regime, however, VDM never reached equilibrium with the SM bath and Eq.~\eqref{eq:Tratio} does not apply. This case requires to solve an unintegrated Boltzmann equation in terms of the energy distribution of $f_\chi$ and evaluate the integral $\rho_\chi = \int_{m_E}^\infty E_\chi f_\chi(E_\chi)$. However, a rough estimate of the VDM energy density at $T\sim\mathcal{O}$(MeV) shows that this is not necessary. Note that VDM particles are dominantly produced from $E$ decays at $T\sim m_E$ with a typical energy of $m_E/2$ (up to a small spread due to the thermal kinetic energy of $E$ in the plasma). Since $\chi$ particles have very weak interactions with the SM bath (and among themselves), they cannot efficiently transfer energy to one another and their energy distribution merely redshifts with expansion. Therefore, at $T\sim\MeV$ we have approximately  
\begin{align} 
	\label{eq:density}
	\rho_\chi({\rm MeV}) 
	\sim \frac{\MeV}{2} n_\chi(\MeV)
	\sim \frac{10^{-3}}{\mchik}\MeV^4\,,
\end{align}
where the second equality assumes $\Omega_\chi h^2=0.12$. Hence,
\begin{align}
	\label{eq:NeffFI}
	\Delta N_{\rm eff}^\chi \sim 2\times 10^{-3}\mchik^{-1}\,.
\end{align} 
which is at most $\Delta N_{\rm eff}^\chi\lesssim 0.03$ given the lower bound $m_\chi\gtrsim 67\,$eV in the freeze-in case.

The above VDM contributions, in Eqs.~\eqref{eq:NeffFO} and~\eqref{eq:NeffFI}, are to be compared to the value $N_{\rm eff} =2.95\pm0.28\,(1\sigma)$~\cite{Cyburt:2015mya} obtained from fitting  helium and deuterium abundance observations within the standard BBN scenario. A value that is consistent with the SM expectation from neutrinos $N_{\rm eff}^\nu=3.046$.  Therefore, we conclude that VDM is not in tension with BBN. \\

Finally, note that while $E$ particles decouple relatively early, they do not decay right away to SM bath particles and $\chi$, due to the small couplings required to explain the DM relic abundance, most notably in the freeze-in regime. Summing the widths into $\chi e$, $Ze$ and $W\nu$ final state in Eqs.~\eqref{E2eXwidth} and~\eqref{E2EWwidth}, the $E$ lifetime is found to be 
\begin{align}
	\tau_E 
	\simeq 2.3\times 10^{-11}\,{\rm sec}\times \frac{\mchik}{\mET^2}\left(1+0.46\frac{\mchik}{\mET}\right)^{-1}
\end{align}
where $\Omega_\chi h^2=0.12$ was imposed, assuming freeze-in production of VDM and $\theta_R=10^{-2}$. Hence, $E$ disappears from the Universe long before BBN starts.  

\paragraph{CMB}          
The $\chi\to 3 \gamma$ decays change the ionization history of the universe and leave imprints on the CMB in the form of observable temperature anisotropies, unless $\tau_{\chi\to3\gamma}\xi^{-1}\gtrsim10^{25}\,\sec$~\cite{Poulin:2016anj} where $\xi\equiv \Omega_\chi/\Omega_{\rm DM}$ denotes the VDM fraction of the observed DM density. As for the constraint from diffuse $\gamma$-rays above, we implicitly assume some late dilution mechanism in the case of the overabundant $\chi$ production. The resulting exclusion, shown in Fig.~\ref{fig:constraints} (magenta shaded region), is slightly weaker than the constraint from diffuse $\gamma$-ray observations, and only relevant for VDM masses close to the electron-positron threshold. 

\paragraph{Structure formation} 
Once produced, DM particles propagate freely in the Universe, thus erasing matter density perturbations and, consequently, structures on scales smaller than the free-streaming length $\lamFS$. High-redshift observations, like the Lyman-$\alpha$ forest~\cite{Croft:2000hs,seljak2006cosmological}, forbids strong suppression of the matter power spectrum below $\lamFS\lesssim \cO(0.1\,{\rm Mpc})$~\cite{Viel:2013fqw}.
In order to estimate the resulting bound on the VDM parameter space, we follow  Ref.~\cite{Falkowski:2017uya} and simply demand that the VDM free-streaming scale does not exceed $0.1\,{\rm Mpc}$.~\footnote{Another approach, giving comparable estimates, consists in mapping analytically models of nonthermal DM to a warm DM (WDM) candidate and recast existent WDM analyses of the matter power spectrum~\cite{Kamada:2019kpe,Huo:2019bjf}.} 

In the radiation-dominated era, the DM free-streaming scale is~\cite{Kolb:1990vq} 
\begin{align}
	\lamFS
	\approx
	\frac{a_{\rm NR}}{H_0 \sqrt{\Omega_R}}
	\left[
	0.62 + \log\left( \frac{a_{\rm eq}}{a_{\rm NR}} \right)
	\right] \, ,
\end{align}
where $a_{\rm eq}=2.9 \times 10^{-4}$ is the scale factor of the Universe at the time of matter/radiation equality, $\Omega_R=9.3\times 10^{-5}$ is the radiation density and $H_0=67.3 \,{\rm km} \,{\rm s}^{-1}\,{\rm Mpc}^{-1}$ is the Hubble's constant today. Numerical values of these cosmological parameters are taken from the results of the full-mission Planck measurements~\cite{Ade:2015xua}. $a_{\rm NR}$ denotes the scale factor when the VDM became non-relativistic. The fact that momentum redshifts like $a^{-1}$ allows to relate to a different time, for instance the VDM decoupling time $t_d$, with    
\begin{align}
	\label{eq:aNR}
	a_{\rm NR}
=	\frac{\avg{p}_{d}}{\avg{p}_{\rm NR}} a_{d} \,
\end{align}
where $\avg{p}$ is the averaged VDM momemtum, with $\avg{p}_{\rm NR}= m_\chi$. The scale factor $a_d$ is related to its value today by entrody dilution $a_d=a_0[g_s(T_0)/g_s(T_d)]^{1/3}(T_0/T_d)$, where $T=T_0\approx 2.3\times 10^{-4}\,$eV and  $a_0=1$, conventionally.\\

In the freeze-out  limit, VDM decouples at $T_d=T_f$ with a thermal distribution, giving $a_{\rm NR}\simeq 3.15T_0/m_\chi[g_s(T_0)/g_s(T_f)]^{1/3}$   and 
\begin{align}
	\lambda_{FS}
=	\frac{0.86\,{\rm Mpc}}{\mchik}\left(1+0.13\log \mchik\right)\,,
\end{align}
taking $g_s(T_0)\approx 3.91$ and $g_s(T_f)\approx 106.5$. Forbidding VDM to stream freely on distances larger than $\sim0.1\,$Mpc implies a lower bound on the VDM mass of  $m_\chi \gtrsim 11\,$keV. This scenario is thus excluded by structure formation since the correct relic density in the freeze-out limit requires $m_\chi\sim60\,$eV. 

The estimation of the free-streaming length differs in the freeze-in case since VDM never thermalized. Instead, most of it is produced at $T_d\sim m_E/3$ with $\avg{p}_d\sim m_E/2$. Consequently, $a_{\rm NR}\simeq 3T_0/(2m_\chi)[g_s(T_0)/g_s(T_d)]^{1/3}$ and 
\begin{align}
	\lambda_{FS}
=	\frac{0.45\,{\rm Mpc}}{\mchik}\left(1+0.12\log \mchik\right)\,,
\end{align}
yielding a lower bound on the VDM mass of 
\begin{align}
	m_\chi\gtrsim 5.4\,{\rm keV}\,.
\end{align}
The above bounds are shown in Fig.~\ref{fig:constraints} (gray shaded region) assuming all DM is in the form of VDM. In the region where the freeze-in production is overefficient the VDM relic density is implicitly assumed to be diluted at late times to the observed value. For the case that VDM is underabundant, a somewhat weaker constraint is expected~\cite{Boyarsky:2008xj}.

\paragraph{Stellar cooling}
Light particles of sub-MeV mass can be produced inside the hot and dense interior of stars. 
If weakly coupled to particles in the plasma, they will propagate without rescattering and eventually escape.
Hence, the  light particles produced will carry away energy and contribute an additional mechanism for stellar cooling, which is constrained by stellar modeling and observations~\cite{Redondo:2008aa, An:2013yfc,Redondo:2013lna}. Moreover, if the new particle mass is close to the plasma frequency in the stellar medium,  emission happens resonantly which substantially enhances energy losses.

VDM particles produced within stellar cores follow from two processes:  bremsstrahlung and inverse Compton scattering, both of which are sensitive to the VDM coupling to electron. Here we recast the bounds of Refs.~\cite{An:2013yfc,Redondo:2013lna,An:2014twa} obtained for dark photons using the Sun, Horizontal Branch~(HB) stars and Red Giants~(RGs), by requiring that VDM emission be at most 10\,\% of the observed luminosity for the first two, and smaller than 10/erg/g/s for the last one. As for dark photons, the production of spin-1 particles is dominated by mixing with the photon in the stellar medium.  This can be understood from the fact that, as far as only electrons are concerned which is the case in stars, it is always possible to redefine fields and move to a different basis where the vector coupling of the spin-1 state to electrons is replaced by a kinetic mixing operator $\chi_{\mu\nu}F^{\mu\nu}$, where $F_{\mu\nu}$ denotes the QED field strength. As a result, the cooling mechanism is dominated by the vector part of the electron coupling. 
The resulting bounds are presented in Fig.~\ref{fig:constraints} (green shaded region), where the three dips denote (from left to right) the best sensitivity from the Sun, HB stars and RGs.

%%%%%%%%%%%%%%%%%%%%%%%%%%%%%%%%%%%%%%%%%%%%%%%%%%%%%%%%%%%%%%%%%%%%%%
\subsection{Collider signatures }
%%%%%%%%%%%%%%%%%%%%%%%%%%%%%%%%%%%%%%%%%%%%%%%%%%%%%%%%%%%%%%%%%%%%%%

For $m_E$ in the TeV range, the model studied above has also interesting collider signatures. While the DM itself cannot be detected at colliders, the heavy lepton $E$, which is an important ingredient allowing to reproduce the DM relic density, can be produced at the LHC or future colliders and searched for in multiple ways. 

The dominant $E$ production mechanism is in pairs from EW interactions, namely through $q\bar q$ annihilating to off-shell $Z$ or $\gamma$ in the $s$-channel,
\begin{align}
	q \bar{q} \to  \gamma^\ast / Z^\ast \to  E \overline{ E}  
\end{align}
Using {\sc MadGraph\,5}~v2.7.3~\cite{Alwall:2014hca}, we estimate the $pp\to EE$ cross section at the LHC with $\sqrt{s}=13\,\TeV$ to be $5.4 \times 10^{-2}\,$fb for $m_E=1\,$TeV. Single production of $E$ is also possible, yet at the expense of small mixing angle with the electron, leading to negligible cross sections at colliders. 

Once produced, heavy leptons decay to $\chi e $, $Ze$ or $W\nu$, leading to different final states when $E$ decays promptly, as in the freeze-out regime.   
First, consider the case where $\Gamma_{E\to\chi e}$ dominates over the EW channels, leading to $\chi \chi e^+ e^-$ in the final state. 
This signal is very similar to that of chargino pairs decaying into dilepton plus missing energy~\cite{Khachatryan:2014qwa,ATLAS:2016uwq}. Currently, such searches bound the cross section for chargino pair production at the $\cO(0.1)$fb level~\cite{Aad:2019vnb}. Hence, heavy leptons  with $m_E=1\,\TeV$ or higher are allowed in this case. Then, consider the opposite case where $E$ dominantely decays to EW channels, either $Ze$ or $W\nu$. This leads to clean signatures with multiple charged leptons  in the final state, assuming the $Z$ and $W$ to decay leptonically.  
While a detailed analysis is beyond the scope of the this work, we note that 
current bounds from multilepton searches at the LHC does not reach yet the TeV scale mass region~\cite{Sirunyan:2019bgz,Sirunyan:2019ofn,Aaboud:2018zeb}. For instance, in the $ZZe^+e^-$ channel, taking at least one $Z$ to decay into $e^+e^-$ or $\mu^+\mu^-$, the expected number of multilepton events is less than one  for a data set of  about 140\,fb$^{-1}$ using the above estimate of the $pp\to E\bar E$ cross section at $\sqrt{s}=13\,$TeV. 
Therefore, we conclude that $m_E=1\,TeV$ is allowed by current experimental data.\\ 

For part of the parameter space, in particular when the VDM relic density is produced by freeze-in, $E$ can have finite decay length. 
For instance, imposing $\Omega_\chi h^2=0.12$ yields $c\tau_E\approx 1.5\,{\rm cm}/[(\theta_R/10^{-2})^2+2.2\mchik^{-1}]$ for $m_E=1\,$TeV. 
Thus, LHC searches for long-lived charged particles~\cite{Aaboud:2019trc,Khachatryan:2016sfv,Aaboud:2016uth,ATLAS:2014fka} have some potential to probe TeV-scale heavy leptons that are either stable on collider scale or decay within the detector, notably for small values of the leptonic mixing angle.\\

Finally, precision electroweak observables can be modified in the presence of nonzero values of $\theta_R$ and/or the kinetic mixing parameter $\epsilon$. 
The latter induces a universal shift of the neutral current, whereas the former induces a breaking of lepton-flavor universality of the $Z$ couplings of $\cO(\theta_R^2)$. Moreover, a nonzero kinetic mixing  alters the relation between $m_Z$ and the SM parameters, hence contributing to the $\rho$ parameter, and the $Z$ decay width. 
However, in the limit of $m_\chi\ll m_Z$, the upper bound on $\kinmix$ is  $\cO(10^{-3})$~\cite{Hook:2010tw,Cline:2014dwa,Curtin:2014cca}, which is much larger that the typical loop-induced value considered here. Similarly, deviations from flavor universality in the lepton sector are excluded above the $10^{-3}$ level~\cite{Efrati:2015eaa,Altmannshofer:2014cfa,Falkowski:2019hvp}, corresponding to $\theta_R\lesssim 3\times 10^{-2}$. We conclude that these bounds are too weak to constrain the region of parameter space relevant for DM.

%%%%%%%%%%%%%%%%%%%%%%%%%%%%%%%%%%%%%%%%%%%%%%%%%%%%%%%%%%%%%%%%%%%%%%
\section{Conclusions}
\label{sec:summary}
%%%%%%%%%%%%%%%%%%%%%%%%%%%%%%%%%%%%%%%%%%%%%%%%%%%%%%%%%%%%%%%%%%%%%%

In this work we considered models of decaying spin-1 DM $\chi$ associated with a spontaneously broken U(1)$_X$ gauge symmetry. In constrast with previous models of this kind, the dominant interaction of DM with the SM sector is through a Yukawa portal where the dark-Higgs scalar connects a dark vector-like lepton $E$, charged under U(1)$_X$, to the right-handed electron. In the U(1)$_X$ broken phase, this portal induces a mass mixing between the dark lepton and the electron. There is no stabilizing symmetry for $\chi$. However, for small enough U(1)$_X$ gauge coupling and leptonic mixing angle $\theta_R$, its decays are sufficiently suppressed to guarantee the stability of DM on cosmological scale. The DM relic abundance is set dominantly by $E\to \chi e$ decays in the early Universe. For small portal coupling values such that $\chi$ is never in thermal equilibrium with the SM bath, the DM abundance is produced by the freeze-in mechanism. Conversely, for coupling values larger than $\sim 10^{-7}$, $\chi$ reaches equilibrium before $E$ decouples and the DM abundance is instead set by a relativistic freeze-out mechanism.       
In the latter case, the comoving density of $\chi$ is almost independent of its mass and couplings, and the observed abundance is accomodated only for a specific mass of $m_\chi\sim 60\,\eV$, with a mild dependence (few eVs) on the portal coupling through the effective number of degrees of freedom of the thermal bath at freeze-out. However, this hot DM scenario is excluded by structure formation considerations. 
For higher $\chi$ masses, the correct relic abundance is obtained by the freeze-in mechanism. For $m_\chi> 2m_e\approx1\,$MeV, $\chi$ decays too rapidly into electron pairs to form a valid DM candidate. We stress that successful DM phenomenology is achieved without resorting to ad-hod stabilizing symmetries, nor unnaturally small parameters since  the model has an enhanced symmetry associated with $E$-number conservation in the limit of zero Yukawa portal coupling and gauge couplings always self-renormalize.

Then, we explored different experimental probes of such decaying spin-1 DM, including direct and indirect DM detection, energy losses in stars and the production of diffuse $\gamma$-rays. We also considered implications for BBN and CMB physics, and briefly outlined possible collider signatures in the case that the dark lepton lies around the TeV scale. In conclusion, we find that spin-1 DM in our model can explain the observed relic density while evading all existing constraints. Furthermore, this spin-1 decaying DM would most likely remain invisible also in the next round of DM detection and collider experiments, thus confering on such a DM candidate a stealth character. This feature is readily understood from the parametric relation between the relic density, which is set by $E\to \chi e$ decays whose rate are quadratic in $\theta_R\ll 1$, and the experimental signatures which rely on the DM coupling to electron pairs, suppressed here by higher powers of $\theta_R$. 

We also discussed in our study the impact of having a nonzero kinetic mixing $\kinmix$ with the SM hypercharge. While it does not affect DM production in the early Universe, kinetic mixing does however breaks the latter correlation with DM searches by contributing to the (vector part of the) DM-electron coupling, $g_{\chi ee}^V\sim g_X\theta_R^2/2+\kinmix ec_W$, which results in a stronger DM detection signals. Considering for illustration $\epsilon=g_Xg_Y/(6\pi^2)$, which is commensurate with the typical $E$ contribution at one-loop, the enhanced signals are still below current bounds for $\theta_R\sim\cO(10^{-2}-10^{-4})$. However, for smaller mixing angle, including values such that the relic density is dominantly set by scattering processes, only part of the observed DM can be explained within our model.    All our results are summarized in Fig.~\ref{fig:constraints}. \\

The model presented in this work admits various modifications, which can give different phenomenology.
For instance, the new lepton can mix with $\mu$ and/or $\tau$, instead of the electron. While the relic density can be accomodated in a similar way, with the possibility to raise the DM mass up to the $2\mu$ or $2m_\tau$ threshold, this would however tear down hopes of direct DM detection, as these experiments rely on the DM coupling to electrons. 
Another possibility is mixing with the quark sector by introducing dark vector-like quarks, instead of leptons.  
Such models would bring about a distinct and rich phenomenology, most notably in terms of direct detection since DM could be at the GeV mass scale and couple to nucleons. Moreover, through CKM mixing effects dark sector contributions to flavor violating processes are unavoidable. Similarly, in leptophilic models, lepton-flavor violating processes such as $\mu\to e\gamma$ could be relevant, if vector-like leptons mix with more than one SM lepton flavor. However, such effects are expected be negligible due to the small U(1)$_X$ gauge coupling needed for a valid DM candicate.

%%%%%%%%%%%%%%%%%%%%%%%%%%%%%%%%%%%%%%%%
\section*{Acknowledgements}
%%%%%%%%%%%%%%%%%%%%%%%%%%%%%%%%%%%%%%%%

We thank Yi Fan Chen, Shlomit Tarem and Tomer Volansky for fruitful discussions and Genevi\`eve B\'elanger, Eric Kuflik and Yue Zhao for comments on the manuscript. 
TM is supported by the Israel Science Foundation (Grant No. 751/19).
TM and YS are supported by the United States-Israel Binational Science Foundation~(BSF) (NSF-BSF program Grant No. 2018683) and the Azrieli foundation.
YS is Taub fellow (supported by the Taub Family Foundation). 

%%%%%%%%%%%%%%%%%%%%%%%%%%%%%%%%%%%%%%%%%%%%%%%%%%%%%%%%%%%%%%%%%%%%%%

%%%%%%%%%%%%%%%%%%%%%%%%%%%%%%%%%%%%%%%%%%%%%%%%%%%%%%%%%%%%%%%%%%%%%%
\bibliographystyle{JHEP}
\bibliography{references}

\providecommand{\href}[2]{#2}\begingroup\raggedright\begin{thebibliography}{10}

\bibitem{PhysRevD.31.3059}
M.~W. Goodman and E.~Witten, \emph{Detectability of certain dark-matter
  candidates}, \href{https://doi.org/10.1103/PhysRevD.31.3059}{\emph{Phys. Rev.
  D} {\bfseries 31} (Jun, 1985) 3059--3063}.

\bibitem{Akerib:2016vxi}
{\scshape LUX} collaboration, D.~S. Akerib et~al., \emph{{Results from a search
  for dark matter in the complete LUX exposure}},
  \href{https://doi.org/10.1103/PhysRevLett.118.021303}{\emph{Phys. Rev. Lett.}
  {\bfseries 118} (2017) 021303},
  [\href{https://arxiv.org/abs/1608.07648}{{\ttfamily 1608.07648}}].

\bibitem{Cui:2017nnn}
{\scshape PandaX-II} collaboration, X.~Cui et~al., \emph{{Dark Matter Results
  From 54-Ton-Day Exposure of PandaX-II Experiment}},
  \href{https://doi.org/10.1103/PhysRevLett.119.181302}{\emph{Phys. Rev. Lett.}
  {\bfseries 119} (2017) 181302},
  [\href{https://arxiv.org/abs/1708.06917}{{\ttfamily 1708.06917}}].

\bibitem{Aprile:2018dbl}
{\scshape XENON} collaboration, E.~Aprile et~al., \emph{{Dark Matter Search
  Results from a One Ton-Year Exposure of XENON1T}},
  \href{https://doi.org/10.1103/PhysRevLett.121.111302}{\emph{Phys. Rev. Lett.}
  {\bfseries 121} (2018) 111302},
  [\href{https://arxiv.org/abs/1805.12562}{{\ttfamily 1805.12562}}].

\bibitem{Du:2018uak}
{\scshape ADMX} collaboration, N.~Du et~al., \emph{{A Search for Invisible
  Axion Dark Matter with the Axion Dark Matter Experiment}},
  \href{https://doi.org/10.1103/PhysRevLett.120.151301}{\emph{Phys. Rev. Lett.}
  {\bfseries 120} (2018) 151301},
  [\href{https://arxiv.org/abs/1804.05750}{{\ttfamily 1804.05750}}].

\bibitem{Ouellet:2018beu}
J.~L. Ouellet et~al., \emph{{First Results from ABRACADABRA-10 cm: A Search for
  Sub-$\mu$eV Axion Dark Matter}},
  \href{https://doi.org/10.1103/PhysRevLett.122.121802}{\emph{Phys. Rev. Lett.}
  {\bfseries 122} (2019) 121802},
  [\href{https://arxiv.org/abs/1810.12257}{{\ttfamily 1810.12257}}].

\bibitem{Battaglieri:2017aum}
M.~Battaglieri et~al., \emph{{US Cosmic Visions: New Ideas in Dark Matter 2017:
  Community Report}},  in \emph{{U.S. Cosmic Visions: New Ideas in Dark Matter
  College Park, MD, USA, March 23-25, 2017}}, 2017,
  \href{https://arxiv.org/abs/1707.04591}{{\ttfamily 1707.04591}},
  \href{http://lss.fnal.gov/archive/2017/conf/fermilab-conf-17-282-ae-ppd-t.pdf}{http://lss.fnal.gov/archive/2017/conf/fermilab-conf-17-282-ae-ppd-t.pdf}.

\bibitem{Garny:2015sjg}
M.~Garny, M.~Sandora and M.~S. Sloth, \emph{{Planckian Interacting Massive
  Particles as Dark Matter}},
  \href{https://doi.org/10.1103/PhysRevLett.116.101302}{\emph{Phys. Rev. Lett.}
  {\bfseries 116} (2016) 101302},
  [\href{https://arxiv.org/abs/1511.03278}{{\ttfamily 1511.03278}}].

\bibitem{Garny:2017kha}
M.~Garny, A.~Palessandro, M.~Sandora and M.~S. Sloth, \emph{{Theory and
  Phenomenology of Planckian Interacting Massive Particles as Dark Matter}},
  \href{https://doi.org/10.1088/1475-7516/2018/02/027}{\emph{JCAP} {\bfseries
  1802} (2018) 027}, [\href{https://arxiv.org/abs/1709.09688}{{\ttfamily
  1709.09688}}].

\bibitem{Ema:2018ucl}
Y.~Ema, K.~Nakayama and Y.~Tang, \emph{{Production of Purely Gravitational Dark
  Matter}}, \href{https://doi.org/10.1007/JHEP09(2018)135}{\emph{JHEP}
  {\bfseries 09} (2018) 135},
  [\href{https://arxiv.org/abs/1804.07471}{{\ttfamily 1804.07471}}].

\bibitem{Garny:2018grs}
M.~Garny, A.~Palessandro, M.~Sandora and M.~S. Sloth, \emph{{Charged Planckian
  Interacting Dark Matter}},
  \href{https://doi.org/10.1088/1475-7516/2019/01/021}{\emph{JCAP} {\bfseries
  1901} (2019) 021}, [\href{https://arxiv.org/abs/1810.01428}{{\ttfamily
  1810.01428}}].

\bibitem{McDonald:2001vt}
J.~McDonald, \emph{{Thermally generated gauge singlet scalars as
  selfinteracting dark matter}},
  \href{https://doi.org/10.1103/PhysRevLett.88.091304}{\emph{Phys. Rev. Lett.}
  {\bfseries 88} (2002) 091304},
  [\href{https://arxiv.org/abs/hep-ph/0106249}{{\ttfamily hep-ph/0106249}}].

\bibitem{Hall:2009bx}
L.~J. Hall, K.~Jedamzik, J.~March-Russell and S.~M. West, \emph{{Freeze-In
  Production of FIMP Dark Matter}},
  \href{https://doi.org/10.1007/JHEP03(2010)080}{\emph{JHEP} {\bfseries 03}
  (2010) 080}, [\href{https://arxiv.org/abs/0911.1120}{{\ttfamily 0911.1120}}].

\bibitem{Bernal:2017kxu}
N.~Bernal, M.~Heikinheimo, T.~Tenkanen, K.~Tuominen and V.~Vaskonen, \emph{{The
  Dawn of FIMP Dark Matter: A Review of Models and Constraints}},
  \href{https://doi.org/10.1142/S0217751X1730023X}{\emph{Int. J. Mod. Phys.}
  {\bfseries A32} (2017) 1730023},
  [\href{https://arxiv.org/abs/1706.07442}{{\ttfamily 1706.07442}}].

\bibitem{Elahi:2014fsa}
F.~Elahi, C.~Kolda and J.~Unwin, \emph{{UltraViolet Freeze-in}},
  \href{https://doi.org/10.1007/JHEP03(2015)048}{\emph{JHEP} {\bfseries 03}
  (2015) 048}, [\href{https://arxiv.org/abs/1410.6157}{{\ttfamily 1410.6157}}].

\bibitem{Falkowski:2017uya}
A.~Falkowski, E.~Kuflik, N.~Levi and T.~Volansky, \emph{{Light Dark Matter from
  Leptogenesis}}, \href{https://doi.org/10.1103/PhysRevD.99.015022}{\emph{Phys.
  Rev.} {\bfseries D99} (2019) 015022},
  [\href{https://arxiv.org/abs/1712.07652}{{\ttfamily 1712.07652}}].

\bibitem{Heeba:2018wtf}
S.~Heeba, F.~Kahlhoefer and P.~St\"ocker, \emph{{Freeze-in production of
  decaying dark matter in five steps}},
  \href{https://doi.org/10.1088/1475-7516/2018/11/048}{\emph{JCAP} {\bfseries
  11} (2018) 048}, [\href{https://arxiv.org/abs/1809.04849}{{\ttfamily
  1809.04849}}].

\bibitem{Fradette:2018hhl}
A.~Fradette, M.~Pospelov, J.~Pradler and A.~Ritz, \emph{{Cosmological beam
  dump: constraints on dark scalars mixed with the Higgs boson}},
  \href{https://doi.org/10.1103/PhysRevD.99.075004}{\emph{Phys. Rev. D}
  {\bfseries 99} (2019) 075004},
  [\href{https://arxiv.org/abs/1812.07585}{{\ttfamily 1812.07585}}].

\bibitem{Pospelov:2008jk}
M.~Pospelov, A.~Ritz and M.~B. Voloshin, \emph{{Bosonic super-WIMPs as
  keV-scale dark matter}},
  \href{https://doi.org/10.1103/PhysRevD.78.115012}{\emph{Phys. Rev.}
  {\bfseries D78} (2008) 115012},
  [\href{https://arxiv.org/abs/0807.3279}{{\ttfamily 0807.3279}}].

\bibitem{Redondo:2008ec}
J.~Redondo and M.~Postma, \emph{{Massive hidden photons as lukewarm dark
  matter}}, \href{https://doi.org/10.1088/1475-7516/2009/02/005}{\emph{JCAP}
  {\bfseries 0902} (2009) 005},
  [\href{https://arxiv.org/abs/0811.0326}{{\ttfamily 0811.0326}}].

\bibitem{Bjorken:2009mm}
J.~D. Bjorken, R.~Essig, P.~Schuster and N.~Toro, \emph{{New Fixed-Target
  Experiments to Search for Dark Gauge Forces}},
  \href{https://doi.org/10.1103/PhysRevD.80.075018}{\emph{Phys. Rev.}
  {\bfseries D80} (2009) 075018},
  [\href{https://arxiv.org/abs/0906.0580}{{\ttfamily 0906.0580}}].

\bibitem{Okun:1982xi}
L.~B. Okun, \emph{{LIMITS OF ELECTRODYNAMICS: PARAPHOTONS?}}, {\emph{Sov. Phys.
  JETP} {\bfseries 56} (1982) 502}.

\bibitem{Galison:1983pa}
P.~Galison and A.~Manohar, \emph{{TWO Z's OR NOT TWO Z's?}},
  \href{https://doi.org/10.1016/0370-2693(84)91161-4}{\emph{Phys. Lett.}
  {\bfseries 136B} (1984) 279--283}.

\bibitem{Holdom:1985ag}
B.~Holdom, \emph{{Two U(1)'s and Epsilon Charge Shifts}},
  \href{https://doi.org/10.1016/0370-2693(86)91377-8}{\emph{Phys. Lett.}
  {\bfseries 166B} (1986) 196--198}.

\bibitem{An:2014twa}
H.~An, M.~Pospelov, J.~Pradler and A.~Ritz, \emph{{Direct Detection Constraints
  on Dark Photon Dark Matter}},
  \href{https://doi.org/10.1016/j.physletb.2015.06.018}{\emph{Phys. Lett.}
  {\bfseries B747} (2015) 331--338},
  [\href{https://arxiv.org/abs/1412.8378}{{\ttfamily 1412.8378}}].

\bibitem{Hochberg:2016sqx}
Y.~Hochberg, T.~Lin and K.~M. Zurek, \emph{{Absorption of light dark matter in
  semiconductors}},
  \href{https://doi.org/10.1103/PhysRevD.95.023013}{\emph{Phys. Rev.}
  {\bfseries D95} (2017) 023013},
  [\href{https://arxiv.org/abs/1608.01994}{{\ttfamily 1608.01994}}].

\bibitem{Redondo:2008aa}
J.~Redondo, \emph{{Helioscope Bounds on Hidden Sector Photons}},
  \href{https://doi.org/10.1088/1475-7516/2008/07/008}{\emph{JCAP} {\bfseries
  07} (2008) 008}, [\href{https://arxiv.org/abs/0801.1527}{{\ttfamily
  0801.1527}}].

\bibitem{An:2013yfc}
H.~An, M.~Pospelov and J.~Pradler, \emph{{New stellar constraints on dark
  photons}}, \href{https://doi.org/10.1016/j.physletb.2013.07.008}{\emph{Phys.
  Lett.} {\bfseries B725} (2013) 190--195},
  [\href{https://arxiv.org/abs/1302.3884}{{\ttfamily 1302.3884}}].

\bibitem{Redondo:2013lna}
J.~Redondo and G.~Raffelt, \emph{{Solar constraints on hidden photons
  re-visited}},
  \href{https://doi.org/10.1088/1475-7516/2013/08/034}{\emph{JCAP} {\bfseries
  1308} (2013) 034}, [\href{https://arxiv.org/abs/1305.2920}{{\ttfamily
  1305.2920}}].

\bibitem{McDermott:2019lch}
S.~D. McDermott and S.~J. Witte, \emph{{The Cosmological Evolution of Light
  Dark Photon Dark Matter}},
  \href{https://arxiv.org/abs/1911.05086}{{\ttfamily 1911.05086}}.

\bibitem{Nelson:2011sf}
A.~E. Nelson and J.~Scholtz, \emph{{Dark Light, Dark Matter and the
  Misalignment Mechanism}},
  \href{https://doi.org/10.1103/PhysRevD.84.103501}{\emph{Phys. Rev.}
  {\bfseries D84} (2011) 103501},
  [\href{https://arxiv.org/abs/1105.2812}{{\ttfamily 1105.2812}}].

\bibitem{Graham:2015rva}
P.~W. Graham, J.~Mardon and S.~Rajendran, \emph{{Vector Dark Matter from
  Inflationary Fluctuations}},
  \href{https://doi.org/10.1103/PhysRevD.93.103520}{\emph{Phys. Rev.}
  {\bfseries D93} (2016) 103520},
  [\href{https://arxiv.org/abs/1504.02102}{{\ttfamily 1504.02102}}].

\bibitem{Agrawal:2018vin}
P.~Agrawal, N.~Kitajima, M.~Reece, T.~Sekiguchi and F.~Takahashi, \emph{{Relic
  Abundance of Dark Photon Dark Matter}},
  \href{https://doi.org/10.1016/j.physletb.2019.135136}{\emph{Phys. Lett.}
  {\bfseries B801} (2020) 135136},
  [\href{https://arxiv.org/abs/1810.07188}{{\ttfamily 1810.07188}}].

\bibitem{Co:2018lka}
R.~T. Co, A.~Pierce, Z.~Zhang and Y.~Zhao, \emph{{Dark Photon Dark Matter
  Produced by Axion Oscillations}},
  \href{https://doi.org/10.1103/PhysRevD.99.075002}{\emph{Phys. Rev.}
  {\bfseries D99} (2019) 075002},
  [\href{https://arxiv.org/abs/1810.07196}{{\ttfamily 1810.07196}}].

\bibitem{Dror:2018pdh}
J.~A. Dror, K.~Harigaya and V.~Narayan, \emph{{Parametric Resonance Production
  of Ultralight Vector Dark Matter}},
  \href{https://doi.org/10.1103/PhysRevD.99.035036}{\emph{Phys. Rev.}
  {\bfseries D99} (2019) 035036},
  [\href{https://arxiv.org/abs/1810.07195}{{\ttfamily 1810.07195}}].

\bibitem{Bastero-Gil:2018uel}
M.~Bastero-Gil, J.~Santiago, L.~Ubaldi and R.~Vega-Morales, \emph{{Vector dark
  matter production at the end of inflation}},
  \href{https://doi.org/10.1088/1475-7516/2019/04/015}{\emph{JCAP} {\bfseries
  1904} (2019) 015}, [\href{https://arxiv.org/abs/1810.07208}{{\ttfamily
  1810.07208}}].

\bibitem{Choi:2020dec}
G.~Choi, T.~T. Yanagida and N.~Yokozaki, \emph{{Dark Photon Dark Matter in the
  minimal $B-L$ Model}},  \href{https://arxiv.org/abs/2008.12180}{{\ttfamily
  2008.12180}}.

\bibitem{DAgnolo:2015ujb}
R.~T. D'Agnolo and J.~T. Ruderman, \emph{{Light Dark Matter from Forbidden
  Channels}}, \href{https://doi.org/10.1103/PhysRevLett.115.061301}{\emph{Phys.
  Rev. Lett.} {\bfseries 115} (2015) 061301},
  [\href{https://arxiv.org/abs/1505.07107}{{\ttfamily 1505.07107}}].

\bibitem{Colombi:1995ze}
S.~Colombi, S.~Dodelson and L.~M. Widrow, \emph{{Large scale structure tests of
  warm dark matter}}, \href{https://doi.org/10.1086/176788}{\emph{Astrophys.
  J.} {\bfseries 458} (1996) 1},
  [\href{https://arxiv.org/abs/astro-ph/9505029}{{\ttfamily
  astro-ph/9505029}}].

\bibitem{Bode:2000gq}
P.~Bode, J.~P. Ostriker and N.~Turok, \emph{{Halo formation in warm dark matter
  models}}, \href{https://doi.org/10.1086/321541}{\emph{Astrophys. J.}
  {\bfseries 556} (2001) 93--107},
  [\href{https://arxiv.org/abs/astro-ph/0010389}{{\ttfamily
  astro-ph/0010389}}].

\bibitem{Viel:2013fqw}
M.~Viel, G.~D. Becker, J.~S. Bolton and M.~G. Haehnelt, \emph{{Warm dark matter
  as a solution to the small scale crisis: New constraints from high redshift
  Lyman-$\alpha$ forest data}},
  \href{https://doi.org/10.1103/PhysRevD.88.043502}{\emph{Phys. Rev. D}
  {\bfseries 88} (2013) 043502},
  [\href{https://arxiv.org/abs/1306.2314}{{\ttfamily 1306.2314}}].

\bibitem{Arcadi:2019lka}
G.~Arcadi, A.~Djouadi and M.~Raidal, \emph{{Dark Matter through the Higgs
  portal}},  \href{https://arxiv.org/abs/1903.03616}{{\ttfamily 1903.03616}}.

\bibitem{Cassel:2009pu}
S.~Cassel, D.~M. Ghilencea and G.~G. Ross, \emph{{Electroweak and Dark Matter
  Constraints on a Z-prime in Models with a Hidden Valley}},
  \href{https://doi.org/10.1016/j.nuclphysb.2009.10.029}{\emph{Nucl. Phys.}
  {\bfseries B827} (2010) 256--280},
  [\href{https://arxiv.org/abs/0903.1118}{{\ttfamily 0903.1118}}].

\bibitem{Cline:2014dwa}
J.~M. Cline, G.~Dupuis, Z.~Liu and W.~Xue, \emph{{The windows for kinetically
  mixed Z'-mediated dark matter and the galactic center gamma ray excess}},
  \href{https://doi.org/10.1007/JHEP08(2014)131}{\emph{JHEP} {\bfseries 08}
  (2014) 131}, [\href{https://arxiv.org/abs/1405.7691}{{\ttfamily 1405.7691}}].

\bibitem{PhysRev.77.242}
C.~N. Yang, \emph{Selection rules for the dematerialization of a particle into
  two photons}, \href{https://doi.org/10.1103/PhysRev.77.242}{\emph{Phys. Rev.}
  {\bfseries 77} (Jan, 1950) 242--245}.

\bibitem{Heisenberg:1935qt}
W.~Heisenberg and H.~Euler, \emph{{Consequences of Dirac's theory of
  positrons}}, \href{https://doi.org/10.1007/BF01343663}{\emph{Z. Phys.}
  {\bfseries 98} (1936) 714--732},
  [\href{https://arxiv.org/abs/physics/0605038}{{\ttfamily physics/0605038}}].

\bibitem{McDermott:2017qcg}
S.~D. McDermott, H.~H. Patel and H.~Ramani, \emph{{Dark Photon Decay Beyond The
  Euler-Heisenberg Limit}},
  \href{https://doi.org/10.1103/PhysRevD.97.073005}{\emph{Phys. Rev. D}
  {\bfseries 97} (2018) 073005},
  [\href{https://arxiv.org/abs/1705.00619}{{\ttfamily 1705.00619}}].

\bibitem{Aghanim:2018eyx}
{\scshape Planck} collaboration, N.~Aghanim et~al., \emph{{Planck 2018 results.
  VI. Cosmological parameters}},
  \href{https://arxiv.org/abs/1807.06209}{{\ttfamily 1807.06209}}.

\bibitem{Belanger:2018ccd}
G.~B\'elanger, F.~Boudjema, A.~Goudelis, A.~Pukhov and B.~Zaldivar,
  \emph{{micrOMEGAs5.0 : Freeze-in}},
  \href{https://doi.org/10.1016/j.cpc.2018.04.027}{\emph{Comput. Phys. Commun.}
  {\bfseries 231} (2018) 173--186},
  [\href{https://arxiv.org/abs/1801.03509}{{\ttfamily 1801.03509}}].

\bibitem{Gondolo:1990dk}
P.~Gondolo and G.~Gelmini, \emph{{Cosmic abundances of stable particles:
  Improved analysis}},
  \href{https://doi.org/10.1016/0550-3213(91)90438-4}{\emph{Nucl. Phys. B}
  {\bfseries 360} (1991) 145--179}.

\bibitem{Weldon:1982aq}
H.~Weldon, \emph{{Covariant Calculations at Finite Temperature: The
  Relativistic Plasma}},
  \href{https://doi.org/10.1103/PhysRevD.26.1394}{\emph{Phys. Rev. D}
  {\bfseries 26} (1982) 1394}.

\bibitem{Angle:2011th}
{\scshape XENON10} collaboration, J.~Angle et~al., \emph{{A search for light
  dark matter in XENON10 data}},
  \href{https://doi.org/10.1103/PhysRevLett.110.249901,
  10.1103/PhysRevLett.107.051301}{\emph{Phys. Rev. Lett.} {\bfseries 107}
  (2011) 051301}, [\href{https://arxiv.org/abs/1104.3088}{{\ttfamily
  1104.3088}}].

\bibitem{Aprile:2020tmw}
{\scshape XENON} collaboration, E.~Aprile et~al., \emph{{Observation of Excess
  Electronic Recoil Events in XENON1T}},
  \href{https://arxiv.org/abs/2006.09721}{{\ttfamily 2006.09721}}.

\bibitem{Aprile:2019xxb}
{\scshape XENON} collaboration, E.~Aprile et~al., \emph{{Light Dark Matter
  Search with Ionization Signals in XENON1T}},
  \href{https://doi.org/10.1103/PhysRevLett.123.251801}{\emph{Phys. Rev. Lett.}
  {\bfseries 123} (2019) 251801},
  [\href{https://arxiv.org/abs/1907.11485}{{\ttfamily 1907.11485}}].

\bibitem{Poulin:2016anj}
V.~Poulin, J.~Lesgourgues and P.~D. Serpico, \emph{{Cosmological constraints on
  exotic injection of electromagnetic energy}},
  \href{https://doi.org/10.1088/1475-7516/2017/03/043}{\emph{JCAP} {\bfseries
  03} (2017) 043}, [\href{https://arxiv.org/abs/1610.10051}{{\ttfamily
  1610.10051}}].

\bibitem{Yuksel:2007dr}
H.~Yuksel and M.~D. Kistler, \emph{{Circumscribing late dark matter decays
  model independently}},
  \href{https://doi.org/10.1103/PhysRevD.78.023502}{\emph{Phys. Rev.}
  {\bfseries D78} (2008) 023502},
  [\href{https://arxiv.org/abs/0711.2906}{{\ttfamily 0711.2906}}].

\bibitem{Bloch:2016sjj}
I.~M. Bloch, R.~Essig, K.~Tobioka, T.~Volansky and T.-T. Yu, \emph{{Searching
  for Dark Absorption with Direct Detection Experiments}},
  \href{https://doi.org/10.1007/JHEP06(2017)087}{\emph{JHEP} {\bfseries 06}
  (2017) 087}, [\href{https://arxiv.org/abs/1608.02123}{{\ttfamily
  1608.02123}}].

\bibitem{Essig:2011nj}
R.~Essig, J.~Mardon and T.~Volansky, \emph{{Direct Detection of Sub-GeV Dark
  Matter}}, \href{https://doi.org/10.1103/PhysRevD.85.076007}{\emph{Phys. Rev.}
  {\bfseries D85} (2012) 076007},
  [\href{https://arxiv.org/abs/1108.5383}{{\ttfamily 1108.5383}}].

\bibitem{Barak:2020fql}
{\scshape SENSEI} collaboration, L.~Barak et~al., \emph{{SENSEI:
  Direct-Detection Results on sub-GeV Dark Matter from a New Skipper-CCD}},
  \href{https://arxiv.org/abs/2004.11378}{{\ttfamily 2004.11378}}.

\bibitem{Aprile:2014eoa}
{\scshape XENON100} collaboration, E.~Aprile et~al., \emph{{First Axion Results
  from the XENON100 Experiment}},
  \href{https://doi.org/10.1103/PhysRevD.90.062009,
  10.1103/PhysRevD.95.029904}{\emph{Phys. Rev.} {\bfseries D90} (2014) 062009},
  [\href{https://arxiv.org/abs/1404.1455}{{\ttfamily 1404.1455}}].

\bibitem{Aguilar-Arevalo:2016zop}
{\scshape DAMIC} collaboration, A.~Aguilar-Arevalo et~al., \emph{{First
  Direct-Detection Constraints on eV-Scale Hidden-Photon Dark Matter with DAMIC
  at SNOLAB}},
  \href{https://doi.org/10.1103/PhysRevLett.118.141803}{\emph{Phys. Rev. Lett.}
  {\bfseries 118} (2017) 141803},
  [\href{https://arxiv.org/abs/1611.03066}{{\ttfamily 1611.03066}}].

\bibitem{Tiffenberg:2017aac}
{\scshape SENSEI} collaboration, J.~Tiffenberg, M.~Sofo-Haro, A.~Drlica-Wagner,
  R.~Essig, Y.~Guardincerri, S.~Holland et~al., \emph{{Single-electron and
  single-photon sensitivity with a silicon Skipper CCD}},
  \href{https://doi.org/10.1103/PhysRevLett.119.131802}{\emph{Phys. Rev. Lett.}
  {\bfseries 119} (2017) 131802},
  [\href{https://arxiv.org/abs/1706.00028}{{\ttfamily 1706.00028}}].

\bibitem{Abramoff:2019dfb}
{\scshape SENSEI} collaboration, O.~Abramoff et~al., \emph{{SENSEI:
  Direct-Detection Constraints on Sub-GeV Dark Matter from a Shallow
  Underground Run Using a Prototype Skipper-CCD}},
  \href{https://doi.org/10.1103/PhysRevLett.122.161801}{\emph{Phys. Rev. Lett.}
  {\bfseries 122} (2019) 161801},
  [\href{https://arxiv.org/abs/1901.10478}{{\ttfamily 1901.10478}}].

\bibitem{An:2013yua}
H.~An, M.~Pospelov and J.~Pradler, \emph{{Dark Matter Detectors as Dark Photon
  Helioscopes}},
  \href{https://doi.org/10.1103/PhysRevLett.111.041302}{\emph{Phys. Rev. Lett.}
  {\bfseries 111} (2013) 041302},
  [\href{https://arxiv.org/abs/1304.3461}{{\ttfamily 1304.3461}}].

\bibitem{An:2020bxd}
H.~An, M.~Pospelov, J.~Pradler and A.~Ritz, \emph{{New limits on dark photons
  from solar emission and keV scale dark matter}},
  \href{https://arxiv.org/abs/2006.13929}{{\ttfamily 2006.13929}}.

\bibitem{Shvartsman:1969mm}
V.~F. Shvartsman, \emph{{Density of relict particles with zero rest mass in the
  universe}}, {\emph{Pisma Zh. Eksp. Teor. Fiz.} {\bfseries 9} (1969)
  315--317}.

\bibitem{Steigman:1977kc}
G.~Steigman, D.~N. Schramm and J.~E. Gunn, \emph{{Cosmological Limits to the
  Number of Massive Leptons}},
  \href{https://doi.org/10.1016/0370-2693(77)90176-9}{\emph{Phys. Lett.}
  {\bfseries 66B} (1977) 202--204}.

\bibitem{Sarkar:1995dd}
S.~Sarkar, \emph{{Big bang nucleosynthesis and physics beyond the standard
  model}}, \href{https://doi.org/10.1088/0034-4885/59/12/001}{\emph{Rept. Prog.
  Phys.} {\bfseries 59} (1996) 1493--1610},
  [\href{https://arxiv.org/abs/hep-ph/9602260}{{\ttfamily hep-ph/9602260}}].

\bibitem{Olive:1980wz}
K.~A. Olive, D.~N. Schramm and G.~Steigman, \emph{{Limits on New Superweakly
  Interacting Particles from Primordial Nucleosynthesis}},
  \href{https://doi.org/10.1016/0550-3213(81)90065-1}{\emph{Nucl. Phys. B}
  {\bfseries 180} (1981) 497--515}.

\bibitem{Cyburt:2015mya}
R.~H. Cyburt, B.~D. Fields, K.~A. Olive and T.-H. Yeh, \emph{{Big Bang
  Nucleosynthesis: 2015}},
  \href{https://doi.org/10.1103/RevModPhys.88.015004}{\emph{Rev. Mod. Phys.}
  {\bfseries 88} (2016) 015004},
  [\href{https://arxiv.org/abs/1505.01076}{{\ttfamily 1505.01076}}].

\bibitem{Croft:2000hs}
R.~A. Croft, D.~H. Weinberg, M.~Bolte, S.~Burles, L.~Hernquist, N.~Katz et~al.,
  \emph{{Towards a precise measurement of matter clustering: Lyman alpha forest
  data at redshifts 2-4}},
  \href{https://doi.org/10.1086/344099}{\emph{Astrophys. J.} {\bfseries 581}
  (2002) 20--52}, [\href{https://arxiv.org/abs/astro-ph/0012324}{{\ttfamily
  astro-ph/0012324}}].

\bibitem{seljak2006cosmological}
U.~Seljak, A.~Slosar and P.~McDonald, \emph{Cosmological parameters from
  combining the lyman-$\alpha$ forest with cmb, galaxy clustering and sn
  constraints}, {\emph{Journal of Cosmology and Astroparticle Physics}
  {\bfseries 2006} (2006) 014}.

\bibitem{Kamada:2019kpe}
A.~Kamada and K.~Yanagi, \emph{{Constraining FIMP from the structure formation
  of the Universe: analytic mapping from $m_{\mathrm{WDM}}$}},
  \href{https://doi.org/10.1088/1475-7516/2019/11/029}{\emph{JCAP} {\bfseries
  11} (2019) 029}, [\href{https://arxiv.org/abs/1907.04558}{{\ttfamily
  1907.04558}}].

\bibitem{Huo:2019bjf}
R.~Huo, \emph{{Matter Power Spectrum of Light Freeze-in Dark Matter: With or
  without Self-Interaction}},
  \href{https://doi.org/10.1016/j.physletb.2020.135251}{\emph{Phys. Lett. B}
  {\bfseries 802} (2020) 135251},
  [\href{https://arxiv.org/abs/1907.02454}{{\ttfamily 1907.02454}}].

\bibitem{Kolb:1990vq}
E.~W. Kolb and M.~S. Turner, \emph{{The Early Universe}}, vol.~69.
\newblock 1990.

\bibitem{Ade:2015xua}
{\scshape Planck} collaboration, P.~Ade et~al., \emph{{Planck 2015 results.
  XIII. Cosmological parameters}},
  \href{https://doi.org/10.1051/0004-6361/201525830}{\emph{Astron. Astrophys.}
  {\bfseries 594} (2016) A13},
  [\href{https://arxiv.org/abs/1502.01589}{{\ttfamily 1502.01589}}].

\bibitem{Boyarsky:2008xj}
A.~Boyarsky, J.~Lesgourgues, O.~Ruchayskiy and M.~Viel, \emph{{Lyman-alpha
  constraints on warm and on warm-plus-cold dark matter models}},
  \href{https://doi.org/10.1088/1475-7516/2009/05/012}{\emph{JCAP} {\bfseries
  05} (2009) 012}, [\href{https://arxiv.org/abs/0812.0010}{{\ttfamily
  0812.0010}}].

\bibitem{Alwall:2014hca}
J.~Alwall, R.~Frederix, S.~Frixione, V.~Hirschi, F.~Maltoni, O.~Mattelaer
  et~al., \emph{{The automated computation of tree-level and next-to-leading
  order differential cross sections, and their matching to parton shower
  simulations}}, \href{https://doi.org/10.1007/JHEP07(2014)079}{\emph{JHEP}
  {\bfseries 07} (2014) 079},
  [\href{https://arxiv.org/abs/1405.0301}{{\ttfamily 1405.0301}}].

\bibitem{Khachatryan:2014qwa}
{\scshape CMS} collaboration, V.~Khachatryan et~al., \emph{{Searches for
  electroweak production of charginos, neutralinos, and sleptons decaying to
  leptons and W, Z, and Higgs bosons in pp collisions at 8 TeV}},
  \href{https://doi.org/10.1140/epjc/s10052-014-3036-7}{\emph{Eur. Phys. J.}
  {\bfseries C74} (2014) 3036},
  [\href{https://arxiv.org/abs/1405.7570}{{\ttfamily 1405.7570}}].

\bibitem{ATLAS:2016uwq}
{\scshape ATLAS} collaboration, T.~A. collaboration, \emph{{Search for
  supersymmetry with two and three leptons and missing transverse momentum in
  the final state at $\sqrt{s}=$13\,TeV with the ATLAS detector}}, .

\bibitem{Aad:2019vnb}
{\scshape ATLAS} collaboration, G.~Aad et~al., \emph{{Search for electroweak
  production of charginos and sleptons decaying into final states with two
  leptons and missing transverse momentum in $\sqrt{s}=13$ TeV $pp$ collisions
  using the ATLAS detector}},
  \href{https://doi.org/10.1140/epjc/s10052-019-7594-6}{\emph{Eur. Phys. J.}
  {\bfseries C80} (2020) 123},
  [\href{https://arxiv.org/abs/1908.08215}{{\ttfamily 1908.08215}}].

\bibitem{Sirunyan:2019bgz}
{\scshape CMS} collaboration, A.~M. Sirunyan et~al., \emph{{Search for physics
  beyond the standard model in multilepton final states in proton-proton
  collisions at $\sqrt{s} =$ 13 TeV}},
  \href{https://doi.org/10.1007/JHEP03(2020)051}{\emph{JHEP} {\bfseries 03}
  (2020) 051}, [\href{https://arxiv.org/abs/1911.04968}{{\ttfamily
  1911.04968}}].

\bibitem{Sirunyan:2019ofn}
{\scshape CMS} collaboration, A.~M. Sirunyan et~al., \emph{{Search for
  vector-like leptons in multilepton final states in proton-proton collisions
  at $\sqrt{s}$ = 13 TeV}},
  \href{https://doi.org/10.1103/PhysRevD.100.052003}{\emph{Phys.\ Rev.\ D}
  {\bfseries 100} (2019) 052003},
  [\href{https://arxiv.org/abs/1905.10853}{{\ttfamily 1905.10853}}].

\bibitem{Aaboud:2018zeb}
{\scshape ATLAS} collaboration, M.~Aaboud et~al., \emph{{Search for
  supersymmetry in events with four or more leptons in $\sqrt{s}=13$ TeV $pp$
  collisions with ATLAS}},
  \href{https://doi.org/10.1103/PhysRevD.98.032009}{\emph{Phys.\ Rev.\ D}
  {\bfseries 98} (2018) 032009},
  [\href{https://arxiv.org/abs/1804.03602}{{\ttfamily 1804.03602}}].

\bibitem{Aaboud:2019trc}
{\scshape ATLAS} collaboration, M.~Aaboud et~al., \emph{{Search for heavy
  charged long-lived particles in the ATLAS detector in 36.1 fb$^{-1}$ of
  proton-proton collision data at $\sqrt{s} = 13$ TeV}},
  \href{https://doi.org/10.1103/PhysRevD.99.092007}{\emph{Phys. Rev.}
  {\bfseries D99} (2019) 092007},
  [\href{https://arxiv.org/abs/1902.01636}{{\ttfamily 1902.01636}}].

\bibitem{Khachatryan:2016sfv}
{\scshape CMS} collaboration, V.~Khachatryan et~al., \emph{{Search for
  long-lived charged particles in proton-proton collisions at $\sqrt s=$ 13
  TeV}}, \href{https://doi.org/10.1103/PhysRevD.94.112004}{\emph{Phys. Rev.}
  {\bfseries D94} (2016) 112004},
  [\href{https://arxiv.org/abs/1609.08382}{{\ttfamily 1609.08382}}].

\bibitem{Aaboud:2016uth}
{\scshape ATLAS} collaboration, M.~Aaboud et~al., \emph{{Search for heavy
  long-lived charged $R$-hadrons with the ATLAS detector in 3.2 fb$^{-1}$ of
  proton--proton collision data at $\sqrt{s} = 13$ TeV}},
  \href{https://doi.org/10.1016/j.physletb.2016.07.042}{\emph{Phys. Lett.}
  {\bfseries B760} (2016) 647--665},
  [\href{https://arxiv.org/abs/1606.05129}{{\ttfamily 1606.05129}}].

\bibitem{ATLAS:2014fka}
{\scshape ATLAS} collaboration, G.~Aad et~al., \emph{{Searches for heavy
  long-lived charged particles with the ATLAS detector in proton-proton
  collisions at $ \sqrt{s}=8 $ TeV}},
  \href{https://doi.org/10.1007/JHEP01(2015)068}{\emph{JHEP} {\bfseries 01}
  (2015) 068}, [\href{https://arxiv.org/abs/1411.6795}{{\ttfamily 1411.6795}}].

\bibitem{Hook:2010tw}
A.~Hook, E.~Izaguirre and J.~G. Wacker, \emph{{Model Independent Bounds on
  Kinetic Mixing}}, \href{https://doi.org/10.1155/2011/859762}{\emph{Adv. High
  Energy Phys.} {\bfseries 2011} (2011) 859762},
  [\href{https://arxiv.org/abs/1006.0973}{{\ttfamily 1006.0973}}].

\bibitem{Curtin:2014cca}
D.~Curtin, R.~Essig, S.~Gori and J.~Shelton, \emph{{Illuminating Dark Photons
  with High-Energy Colliders}},
  \href{https://doi.org/10.1007/JHEP02(2015)157}{\emph{JHEP} {\bfseries 02}
  (2015) 157}, [\href{https://arxiv.org/abs/1412.0018}{{\ttfamily 1412.0018}}].

\bibitem{Efrati:2015eaa}
A.~Efrati, A.~Falkowski and Y.~Soreq, \emph{{Electroweak constraints on
  flavorful effective theories}},
  \href{https://doi.org/10.1007/JHEP07(2015)018}{\emph{JHEP} {\bfseries 07}
  (2015) 018}, [\href{https://arxiv.org/abs/1503.07872}{{\ttfamily
  1503.07872}}].

\bibitem{Altmannshofer:2014cfa}
W.~Altmannshofer, S.~Gori, M.~Pospelov and I.~Yavin, \emph{{Quark flavor
  transitions in $L_\mu-L_\tau$ models}},
  \href{https://doi.org/10.1103/PhysRevD.89.095033}{\emph{Phys. Rev.}
  {\bfseries D89} (2014) 095033},
  [\href{https://arxiv.org/abs/1403.1269}{{\ttfamily 1403.1269}}].

\bibitem{Falkowski:2019hvp}
A.~Falkowski and D.~Straub, \emph{{Flavourful SMEFT likelihood for Higgs and
  electroweak data}},
  \href{https://doi.org/10.1007/JHEP04(2020)066}{\emph{JHEP} {\bfseries 04}
  (2020) 066}, [\href{https://arxiv.org/abs/1911.07866}{{\ttfamily
  1911.07866}}].

\end{thebibliography}\endgroup
%%%%%%%%%%%%%%%%%%%%%%%%%%%%%%%%%%%%%%%%%%%%%%%%%%%%%%%%%%%%%%%%%%%%%%

\end{document}